\documentclass[a4paper]{article}

\usepackage[top=25truemm,bottom=25truemm,left=25truemm,right=25truemm]{geometry}

\usepackage{fancyhdr}
\usepackage{fancybox}
\usepackage{ascmac} 
\usepackage{amsmath}
\usepackage{mathtools}
\usepackage{amsfonts}
\usepackage{amssymb}
\usepackage{mathrsfs}
\usepackage{amsthm}
\usepackage{bm} 
\usepackage{enumerate}
\usepackage{lastpage}
\usepackage{graphicx,color}
\usepackage{tikz}
\usetikzlibrary{intersections, calc, arrows}
\usepackage{pgfplots}%
\pgfplotsset{compat=1.12}
\usepackage{version}
\usepackage{comment}
\usepackage{cancel}
\usepackage{url}%
\usepackage{cite}
\allowdisplaybreaks
\usepackage{hyperref}

\newcommand{\drm}{d} 
 
\newcommand{\irm}{i}

\newcommand{\Exp}[1]{\mathrm{e}^{#1}}

\newcommand{\Trm}{\mathrm{T}}

\newcommand{\Rrm}{\mathrm{R}}
\newcommand{\Lrm}{\mathrm{L}}
\newcommand{\Tr}[1]{\mathrm{Tr}{#1}}

\newcommand{\Slash}[1]{{\ooalign{\hfil/\hfil\crcr$#1$}}}

\newcommand{\bra}[1]{\langle #1 |}
\newcommand{\ket}[1]{| #1 \rangle}

\newcommand{\lpartial}{\overset{\leftarrow}{\partial}}

\newcommand{\com}[2]{\left[ #1 , #2 \right]}

\newcommand{\beq}{\begin{eqnarray}}
\newcommand{\eeq}{\end{eqnarray}}

\begin{document}
\begin{flushright}
\today
\end{flushright}
\vspace*{5mm}
\begin{center}
{\large \bf
Twist-3 Gluon Fragmentation Contribution to Hyperon
Polarization\\[10pt]
in Semi-Inclusive Deep Inelastic
Scattering
}
\vspace{1.5cm}\\
{\sc Riku Ikarashi$^1$, Yuji Koike$^2$, 
Kenta Yabe$^1$ and Shinsuke Yoshida$^{3,4}$}
\\[0.7cm]
\vspace*{0.1cm}
{\it $^1$Graduate School of Science and Technology, Niigata University,
Ikarashi 2-no-cho, Niigata 950-2181, Japan}

\vspace{0.2cm}

{\it $^2$Department of Physics, Niigata University, Ikarashi 2-no-cho, Niigata 950-2181, Japan}

\vspace{0.2cm}

{\it $^3$Guangdong Provincial Key Laboratory of Nuclear Science, 
Institute of Quantum Matter,\\
South China Normal University, Guangzhou 510006, China}

\vspace{0.2cm}

{\it $^4$Guangdong-Hong Kong Joint Laboratory of Quantum Matter, 
Southern Nuclear Science Computing Center, 
South China Normal University, Guangzhou 510006, China}
\\[3cm]

{\large \bf Abstract} \end{center}
We derive the twist-3 gluon fragmentation function (FF) contribution to
the transversely polarized hyperon production in semi-inclusive 
deep inelastic scattering, $ep\to e\Lambda^\uparrow X$,  
in the leading order (LO) with respect to the QCD coupling
in the framework of the collinear twist-3 factorization.
Together with the known result for the contribution from the twist-3
distribution in the proton and the twist-3 quark FFs for
the hyperon, this completes the LO cross section for this process.  
The constraint relations among the twist-3 FFs are taken into account.  
The formula is relevant to 
large-$P_T$ hyperon production in the future Electron-Ion-Collider experiment.

\newpage

\section{Introduction}

In a recent paper\cite{Koike:2022ddx}, three of the present authors studied 
the transverse polarization of hyperons
produced in semi-inclusive deep inelastic scattering, $ep\to e\Lambda^\uparrow X$.  
For a large-$P_T$ hyperon production, this process can be analyzed in the framework of the
collinear factorization, in which the
polarization appears as a twist-3 observable in the absence of a leading twist-2 effect.  
For $ep\to e\Lambda^\uparrow X$, the responsible twist-3 effects are
(i) the twist-3 distribution functions (DFs) in the initial proton combined with the twist-2
transversity fragmentation
function (FF) for $\Lambda$ and (ii) the twist-3 FFs for the polarized hyperon 
combined with the twist-2 unpolarized
parton DFs in the proton.  The twist-3 FFs in (ii) are chiral-even, and both
(a) quark and (b) gluon types of twist-3 FFs contribute.  
In \cite{Koike:2022ddx}, the twist-3 polarized cross section for $ep\to e\Lambda^\uparrow X$
from the above (i) and (ii)(a) was derived in the leading order (LO) with respect to the QCD
coupling constant.  As a sequel to \cite{Koike:2022ddx}, we will derive in this paper the 
LO cross section
from (ii)(b), which completes the LO twist-3 cross section for this process.  
Since the gluons are ample in the collision environment and
the twist-3 quark and gluon FFs mix under renormalization, the effect of (ii)(b)
could be as important as (ii)(a).   We also remind that the twist-3 fragmentation effect
is important to understand the single transverse-spin asymmetry
in $p^\uparrow p\to \pi X$\cite{Kanazawa:2014dca,Gamberg:2017gle}, which shows a similar 
rising asymmetry at large $x_F$ as the polarization in $pp\to\Lambda^\uparrow X$.  
Our present study has a direct relevance to
the hyperon polarization phenomenon in the future Electron-Ion-Collider (EIC) experiment.  

Here we make some remarks on the phenomenological use of the twist-3 cross section.  
As we will see, it contains several unknown nonperturbative
functions, determination of which requires global analysis of
data for various processes such as
$ep\to e\Lambda^\uparrow X$, 
$e^+ e^-\to \Lambda^\uparrow X$ and $pp\to \Lambda^\uparrow X$, etc,
combined with an appropriate modelling of those functions.  
We also recall that
in the small-$P_T$ region
the transverse-momentum-dependent (TMD) factorization holds for
$ep\to e\Lambda^\uparrow X$ and 
$e^+ e^-\to \Lambda^\uparrow X$, and 
we anticipate that the two frameworks 
match in the intermediate region of $P_T$\footnote{Study on this matching will be reported elsewhere.}
as for the case of $p^\uparrow p\to\ell^+\ell^- X$\cite{Ji:2006ub} and 
$ep^\uparrow \to e\pi X$
\cite{Ji:2006br,Koike:2007dg,Zhou:2008fb}.  
Information on the TMD functions obtained from the analysis of those small-$P_T$ data will also
help to
constrain the twist-3 functions owing to the
relations between the TMD functions and the twist-3 functions\cite{Kanazawa:2015ajw,Koike:2019zxc}.  
In this connection we mention the
recent data on $e^+ e^-\to \Lambda^\uparrow X$
at Belle\cite{Belle:2018ttu} and the phenomenological analyses of the data
in terms of the TMD factorization\cite{DAlesio:2020wjq,Callos:2020qtu,Chen:2021hdn, Chen:2021zrr}.  
These studies will be useful 
to analyze the EIC data at large-$P_T$ in terms of the twist-3 cross section derived in this work.

The formalism of calculating the twist-3 gluon FFs contribution is very complicated and
was completed only recently for a similar process
in the pp collision, $pp\to\Lambda^\uparrow X$\cite{Koike:2021awj}.  Here we apply the method to
$ep\to e\Lambda^\uparrow X$.  
Since the kinematics for this process was described in \cite{Koike:2022ddx} and
the method is in parallel to the case for $pp\to\Lambda^\uparrow X$\cite{Koike:2021awj},
our presentation in this paper will be brief, referring to those papers for the details.

The remainder of this paper is organized as follows:  In section 2,
we introduce the twist-3 gluon FFs relevant in our study.  In section 3, 
we briefly 
describe the formalism for calculating the 
twist-3 gluon FF contribution to $ep\to e\Lambda^\uparrow X$
and present the LO cross section. 
Section 4 is devoted to a brief summary.

\section{Twist-3 gluon fragmentation functions}

\subsection{Three types of twist-3 gluon FFs and $q\bar{q}g$ FFs}

Here we list the twist-3 gluon FFs for spin-1/2 hyperon which are necessary to
derive the twist-3 cross section for $ep\to e\Lambda^\uparrow X$
\cite{Koike:2019zxc,Koike:2021awj}.
They are classified into the intrinsic, kinematical and dynamical FFs.  
First, the intrinsic gluon FFs are defined as the lightcone correlators of
the gluon's field strength $F^{\mu\nu}_a$ with color index $a$
\cite{Koike:2019zxc,Mulders:2000sh}:
\begin{align}
\label{intrinsic}
 &\widehat\Gamma^{\alpha \beta}(z) = \frac{1}{N^2-1} \sum_{X}
\int \frac{\drm \lambda}{2\pi} \Exp{-\irm \lambda/z}
\bra{0}(\com{\infty w}{0}F^{w \beta}(0))_a
\ket{hX}\bra{hX}
(F^{w \alpha}(\lambda w)\com{\lambda w}{\infty w})_a \ket{0}{\nonumber}\\
 &= -g^{\alpha \beta}_\perp \widehat G(z) 
- i  \epsilon^{P_h w \alpha \beta}(S \cdot w)\Delta \widehat G(z)
+ M_h \epsilon^{P_h w S_\perp \{\alpha}w^{\beta\}}\Delta\widehat G_{3\bar T}(z)
+ i M_h \epsilon^{\alpha \beta  w S_\perp }
\Delta \widehat G_{3T}(z)
+\cdots,  
\end{align}
where $P_h$ is the four momentum of the hyperon with its mass $M_h$.  
$P_h^\mu$ can be regarded as lightlike in the twist-3 accuracy and 
$w^\mu$ is another lightlike vector satisfying $P_h \cdot w=1$.
$S^\mu$ is the spin vector of the hyperon normalized as $S^2=-M_h^2$ and
can be decomposed as
$S^\mu = (S \cdot w) P_h^\mu + (S \cdot P_h)w^\mu + M_h S_\perp^\mu$
with the transverse spin vector $S_\perp^\mu$ ($S_\perp^2=-1$).  
$g^{\alpha\beta}_\perp \equiv g^{\alpha\beta} - P_h^\alpha w^\beta - P_h^\beta w^\alpha$, 
$N=3$ is the number of colors for $SU(N)$ and
the ellipsis denotes twist-4 or higher.
$\ket{h}$ denotes the hyperon state. 
$[\lambda w, \mu w] \equiv \mathcal{P} 
\exp{\left[i g \int_{\mu}^{\lambda} \drm \tau w \cdot A(\tau w)\right]}$ 
is the gauge-link operator which guarantees gauge invariance of the correlation function.
We use the convention for the Levi-Civita symbol as $\epsilon^{0123} =1$.  
The shorthand notation
$\epsilon^{P_h w \alpha \beta} 
\equiv \epsilon^{\mu\nu\alpha\beta}{P_h}_\mu w_\nu$, etc. is used, 
and $\{ \alpha\,\beta \}$ denotes the symmetrization of Lorentz indices.
$\widehat G(z)$ and $\Delta \widehat G(z)$ are 
twist-2 unpolarized and helicity FFs, respectively, and
$\Delta \widehat G_{3 \bar T}(z)$ and $\Delta \widehat G_{3T}(z)$ are intrinsic twist-3 FFs.
All FFs in (\ref{intrinsic}) are defined to be real and have a support on $0<z<1$.
$\Delta \widehat G_{3 \bar T}(z)$ is na\"{i}vely T-odd, and contributes to the hyperon polarization. 

Second, the kinematical gluon FFs are defined from the derivative of the 
correlation functions for the intrinsic one:
\begin{align}
\label{kinematical}
 \widehat \Gamma^{\alpha \beta \gamma}_\partial (z) &=
\frac{1}{N^2 -1}\sum_{X} \int \frac{\drm \lambda}{2 \pi}
\Exp{- \irm \lambda /z} 
\bra{0}(\com{\infty w}{0}F^{w \beta}(0))_a \ket{hX}\bra{hX}
(F^{w \alpha}(\lambda w)\com{\lambda w}{\infty w})_a 
\ket{0}\lpartial{}^\gamma {\nonumber}\\
& =  -\irm \frac{M_h}{2}g_\perp^{\alpha \beta}
\epsilon^{P_h w S_\perp \gamma}\widehat G_T^{(1)}(z)
+ \frac{M_h}{2}\epsilon^{P_h w \alpha \beta}S_\perp^{\gamma}
\Delta \widehat G^{(1)}_T (z)
{\nonumber}\\
 &
  -\irm \frac{M_h}{8}\left(
\epsilon^{P_h w S_\perp \{\alpha}g_\perp^{\beta \} \gamma}
+\epsilon^{P_h w \gamma \{\alpha}S^{\beta \}}_\perp
\right) \Delta \widehat H^{(1)}_T (z) +\cdots, 
\end{align}
where 
\begin{align}
F^{w\alpha}(\lambda w)[\lambda w, \infty w]\ket{0} \lpartial{}^\gamma
\equiv
 \lim_{\xi \to 0} \frac{\drm}{\drm \xi_\gamma}
F^{w\alpha}(\lambda w + \xi)[\lambda w + \xi , \infty w + \xi] \ket{0} .
\end{align}
There are three twist-3 gluonic
kinematical FFs, $\widehat G^{(1)}_T (z)$, $\Delta \widehat G^{(1)}_T (z)$ and $\Delta \widehat H^{(1)}_T (z)$, which 
are real functions and have a support on $0<z<1$.
Among them, $\widehat G^{(1)}_T (z)$ and $\Delta \widehat H^{(1)}_T (z)$
are na\"{i}vely T-odd contributing to the hyperon polarization, while $\Delta \widehat G^{(1)}_T (z)$ is na\"{i}vely T-even.
They can also be written as the $k^2_T / M_h^2$-moment of the TMD FFs 
\cite{Mulders:2000sh}.

Third, the dynamical gluon FFs are defined from the 3-gluon correlators.
Contraction of color indices with
two structure constants for color $SU(N)$, i.e. $- \irm f_{abc}$ and
$d_{abc}$, yields two types of FFs\cite{Koike:2019zxc, Yabe:2019awq, Kenta:2019bxd, Gamberg:2018fwy}:
\begin{align}
\label{dynamicalFA}
& \widehat\Gamma^{\alpha \beta \gamma}_{FA}\left(\frac{1}{z_1},\frac{1}{z_2}\right) {\nonumber}\\
&= \frac{- \irm f_{abc}}{N^2-1}\sum_{X}\iint
\frac{\drm \lambda}{2 \pi}\frac{\drm \mu}{2 \pi}
\Exp{-\irm \lambda /z_1}\Exp{- \irm \mu (1/z_2 - 1/z_1)}
\bra{0}F^{w \beta}_b(0)\ket{hX}
\bra{hX}F^{w \alpha}_a(\lambda w) gF^{w \gamma}_c(\mu w) \ket{0} 
{\nonumber}\\
 &
= -M_h \left(
\widehat N_1\left(\frac{1}{z_1},\frac{1}{z_2}\right)g^{\alpha \gamma}_\perp
\epsilon^{P_h w S_\perp \beta}
+\widehat N_2\left(\frac{1}{z_1},\frac{1}{z_2}\right) g^{\beta \gamma}_\perp
\epsilon^{P_h w S_\perp \alpha} 
  -\widehat N_2\left(\frac{1}{z_2}-\frac{1}{z_1},\frac{1}{z_2}\right)g^{\alpha \beta}_{\perp}
\epsilon^{P_h w S_\perp \gamma}
\right) ,
\end{align}
\begin{align}
\label{dynamicalFS}
 &\widehat\Gamma^{\alpha\beta\gamma}_{FS}\left(\frac{1}{z_1},\frac{1}{z_2}\right) {\nonumber}\\
 &=\frac{d_{abc}}{N^2-1}\sum_{X}\iint\frac{\drm \lambda}{2\pi}
\frac{\drm \mu}{2 \pi}\Exp{-\irm \lambda/z_1}
\Exp{-\irm \mu(1/z_2 - 1/z_1)}
\bra{0}F^{w \beta}_b(0) \ket{hX}\bra{hX}F^{w\alpha}_a(\lambda w) g F^{w\gamma}_c(\mu w) \ket{0}{\nonumber}\\
 &= -M_h \left(\widehat O_1\left(\frac{1}{z_1},\frac{1}{z_2}\right)g^{\alpha \gamma}_\perp
\epsilon^{P_h w S_\perp \beta}+ \widehat O_2\left(\frac{1}{z_1},\frac{1}{z_2}\right)
g^{\beta \gamma}_\perp \epsilon^{P_h w S_\perp \alpha} 
 +\widehat O_2\left(\frac{1}{z_2}-\frac{1}{z_1},\frac{1}{z_2}\right)g^{\alpha \beta}_\perp
\epsilon^{P_h w S_\perp \gamma}\right),
\end{align}
where the gauge-link operators are suppressed for simplicity.  
There are four purely gluonic dynamical FFs,
$\widehat N_{1,2}\left(\frac{1}{z_1},\frac{1}{z_2}\right)$ and 
$\widehat O_{1,2}\left(\frac{1}{z_1}.\frac{1}{z_2}\right)$, 
which are 
complex functions and have a support on $1/z_2>1$ and $1/z_2 > 1/z_1 >0$.
Their real parts 
are na\"{i}vely $T$-even,
while their imaginary parts are na\"{i}vely $T$-odd.
$\widehat N_1\left(\frac{1}{z_1},\frac{1}{z_2}\right)$ and 
$\widehat O_1\left(\frac{1}{z_1},\frac{1}{z_2}\right)$ satisfy the symmetry relations
\begin{align}
\widehat N_1\left(\frac{1}{z_1},\frac{1}{z_2}\right) = 
- \widehat N_1 \left(\frac{1}{z_2} - \frac{1}{z_1}, \frac{1}{z_2}\right), \qquad
\widehat O_1 \left(\frac{1}{z_1},\frac{1}{z_2}\right)= 
\widehat O_1 \left(\frac{1}{z_2}-\frac{1}{z_1},\frac{1}{z_2}\right).
\label{symmetry}
\end{align}

Finally, we introduce other dynamical FFs defined from the quark-antiquark-gluon 
correlators\cite{Koike:2019zxc}, which are necessary for the 
derivation of the twist-3 cross section for $ep \to e\Lambda^\uparrow X$,
\begin{align}
\label{dynamicalDelta}
\widetilde\Delta^\alpha_{ij}\left(\frac{1}{z_1},\frac{1}{z_2}\right)
&=\frac{1}{N}\sum_{X}\iint \frac{\drm \lambda}{2\pi}\frac{\drm \mu}{2\pi}
\Exp{-\irm \lambda/z_1}\Exp{-\irm\mu(1/z_2-1/z_1)}
\bra{0}gF^{w\alpha}_a(\mu w) \ket{hX}\bra{hX}
\bar\psi_j(\lambda w)t^a\psi_i(0)\ket{0}{\nonumber}\\
&= M_h \left(
\epsilon^{\alpha P_h w S_\perp}(\Slash{P}_h)_{ij}
\widetilde D_{FT}\left(\frac{1}{z_1},\frac{1}{z_2}\right)
+ i S^{\alpha }_\perp (\gamma_5 \Slash{P}_h)_{ij}
\widetilde G_{FT}\left(\frac{1}{z_1},\frac{1}{z_2}\right)
\right) ,
\end{align}
where $t^a$ is the generators of SU(N) and the spinor indices $i, j$ are shown explicitly. 
These two functions $\widetilde{D}_{FT}\left(\frac{1}{z_1},\frac{1}{z_2}\right)$ and
$\widetilde{G}_{FT}\left(\frac{1}{z_1},\frac{1}{z_2}\right)$ are 
complex functions and
have a support on $1/z_1 >0,1/z_2 <0$ and $1/z_1 -1/z_2 >1$.
Their real parts are na\"{i}vely $T$-even, while the imaginary parts are na\"{i}vely $T$-odd.
\subsection{Constraint relations among twist-3 gluon FFs}
The gluon FFs introduced above are not independent but are subject to
the QCD equation-of-motion (EOM) relations and the Lorentz invariance relations (LIRs). 
The complete set of those relations were
derived in 
\cite{Koike:2019zxc}.  Here we quote those relations which are useful to simplify
the twist-3 cross section
for $ep\to e\Lambda^\uparrow X$.  
The relevant EOM relation allows us to express the intrinsic FF in terms of the 
kinematical and dynamical FFs as
\begin{align}
\label{eom1}
 &\frac{1}{z}\Delta\widehat G_{3 \bar T}(z) =
- \Im \widetilde D_{FT}(z) + 
\frac{1}{2}\left(\widehat G_T^{(1)}(z)+\Delta\widehat H_T^{(1)}(z) \right)
{\nonumber}\\
 &\qquad+ \int \drm\left(\frac{1}{z'}\right)\frac{1}{1/z -1/z'} \Im
\left(
2 \widehat N_1\left(\frac{1}{z'},\frac{1}{z}\right) + \widehat N_2\left(\frac{1}{z'},\frac{1}{z}\right)-\widehat N_2\left(\frac{1}{z}-\frac{1}{z'},\frac{1}{z}\right)
\right), 
\end{align} 
where $\widetilde D_{FT}(z)$ is defined as
\begin{align}
\widetilde D_{FT}(z) \equiv
\frac{2}{C_F}\int_{0}^{1/z}\drm \left(\frac{1}{z'}\right)
\widetilde D_{FT}\left(\frac{1}{z'},\frac{1}{z'}-\frac{1}{z}\right),  \qquad 
{\rm with}\ C_F=\frac{N^2-1}{2N}.
\label{DFTtild}
\end{align}
Other relations derived from the LIRs and the EOM relations 
represent the derivative of the kinematical FFs in terms of other FFs as
\begin{align}
\label{rel1}
 &\frac{1}{z}\frac{\partial\widehat G_T^{(1)}(z)}{\partial(1/z)}
= -2\left( \Im \widetilde D_{FT}(z) - \widehat G_T^{(1)}(z)\right){\nonumber}\\
 &\qquad+4\int\drm\left(\frac{1}{z'}\right)\frac{1}{1/z-1/z'}
\Im\left(
\widehat N_1\left(\frac{1}{z'},\frac{1}{z}\right) - \widehat N_2\left(\frac{1}{z}-\frac{1}{z'},\frac{1}{z}\right)
\right){\nonumber}\\
&\qquad+2\int\drm\left(\frac{1}{z'}\right)\frac{1/z}{(1/z-1/z')^2}
\Im\left(
\widehat N_1\left(\frac{1}{z'},\frac{1}{z}\right) +\widehat N_2\left(\frac{1}{z'},\frac{1}{z}\right)
-2\widehat N_2\left(\frac{1}{z}-\frac{1}{z'},\frac{1}{z}\right)
\right),
\end{align}
and
\begin{align}
\label{rel2}
& \frac{1}{z}\frac{\partial\Delta\widehat H_T^{(1)}(z)}{\partial(1/z)}
= -4\left(\Im\widetilde D_{FT}(z)-\Delta\widehat H_{T}^{(1)}(z)\right){\nonumber}\\
&\qquad+8\int\drm\left(\frac{1}{z'}\right)\frac{1}{1/z-1/z'}
\Im\left(
\widehat N_1\left(\frac{1}{z'},\frac{1}{z}\right) +\widehat N_2\left(\frac{1}{z'},\frac{1}{z}\right)
\right){\nonumber}\\
&\qquad+4\int\drm \left( \frac{1}{z'} \right)\frac{1/z}{(1/z-1/z')^2}
\Im\left(
\widehat N_1\left(\frac{1}{z'},\frac{1}{z}\right)+\widehat N_2\left(\frac{1}{z'},\frac{1}{z}\right)
\right).
\end{align}
The relations (\ref{eom1}), (\ref{rel1}) and (\ref{rel2}) show that 
the purely gluonic twist-3 FFs are related to the quark-antiquark-gluon 
FFs, which implies the contribution to $ep\to e\Lambda^\uparrow X$
from the latter needs to be considered together.  
It's been shown that the above three relations (\ref{eom1}), (\ref{rel1}) and (\ref{rel2})
are crucial to guarantee the frame independence of the cross section for $pp\to\Lambda^\uparrow X$. 
Using these relations, we will express the cross section in terms of 
$\widehat G^{(1)}_T$, $\Delta \widehat H^{(1)}_T$,
$\Im \widehat N_{1,2}$, $\Im \widehat O_{1,2}$,
$\Im \widetilde{D}_{FT}$ and $\Im \widetilde{G}_{FT}$ (see eq. (\ref{result}) below), which
gives the most concise expression for the cross section.  
We also note that, in principle,
the twist-3 kinematical FFs, $\widehat G^{(1)}_T$ and $\Delta \widehat H^{(1)}_T$, 
can be also eliminated in terms of the twist-3 dynamical FFs
(see eqs. (74) and (75) of \cite{Koike:2019zxc}).

\section{Twist-3 gluon FF contribution to $ep \to e\Lambda^\uparrow X$}
\subsection{Kinematics}
\label{kinematics}

Here we briefly summarize the kinematics for the process\cite{Koike:2022ddx}, 
\beq
e(\ell) + p (p) \to e (\ell') + \Lambda^{\uparrow}(P_h, S_\perp) + X,  
\label{epeLX}
\eeq
where $\ell$, $\ell'$, $p$ and $P_h$ are the momenta of each particle and
$S_\perp$ is the transverse spin vector for $\Lambda$.  With the
virtual photon's momentum $q=\ell-\ell'$, 
we introduce the five Lorentz invariants as
\beq
&&S_{ep} \equiv (p+\ell)^2 \simeq 2 p \cdot \ell,\qquad
Q^2 \equiv -q^2,
\nonumber\\
&&x_{bj} \equiv \frac{Q^2}{2 p \cdot q} , \qquad 
z_f \equiv \frac{p \cdot P_h}{p \cdot q},\qquad
q_T \equiv \sqrt{-q_t^2},
\eeq
where
\beq
q_t^\mu \equiv q^\mu - \frac{P_h \cdot q}{p \cdot P_h}p^\mu
-\frac{p \cdot q}{p \cdot P_h} P_h^\mu
\eeq
is a space-like momentum satisfying $q_t \cdot p = q_t \cdot P_h = 0$.
\begin{figure}[h]
 \begin{center}
\includegraphics[scale=0.9]{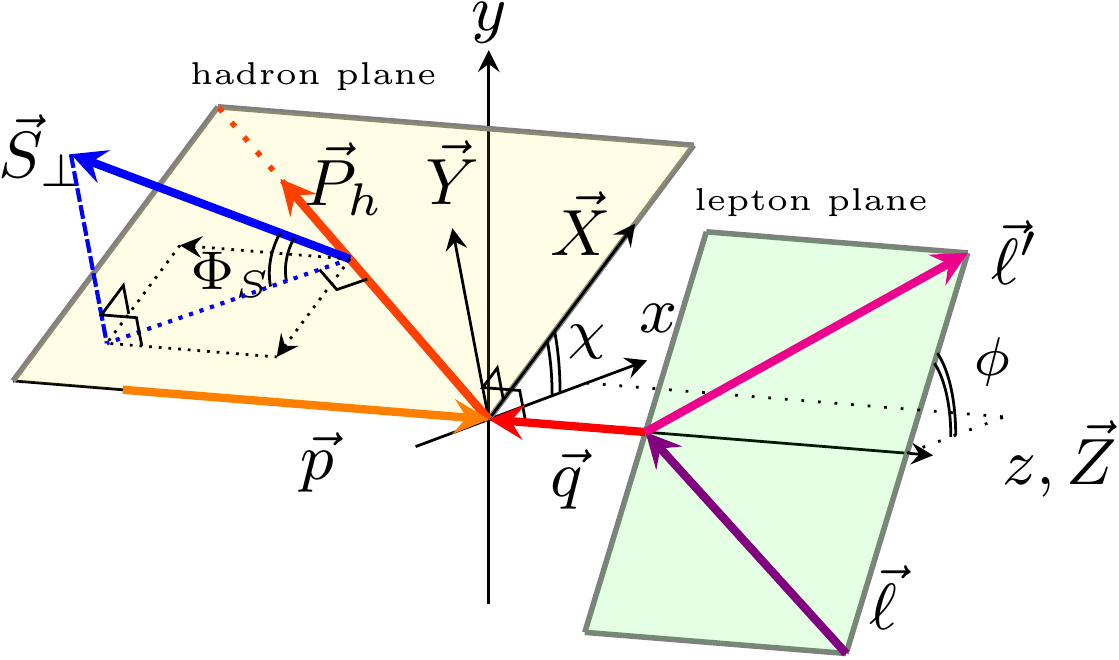}
\vspace{0.3cm}
\caption{Hadron frame and the transverse spin vector $\vec{S}_\perp$.
To make clear the convention for
$\Phi_S$, rotate the $Z$ and $X$ axes around the $Y$ axis by 
$\theta$ (polar angle of $\vec{P}_h$)
so that the new $Z$ axis becomes parallel to $\vec{P}_h$.  $\Phi_S$ is 
defined to be
the azimuthal angle
of $\vec{S}_\perp$ around $\vec{P}_h$ measured from the new $X$ axis,
just like $\phi$ and $\chi$ are measured from the $x$ axis around the $z$ axis.}
\label{hadronframe}
\end{center}
\end{figure}
As in \cite{Koike:2022ddx}, we work in the hadron 
frame\cite{Meng:1991da} (See Fig. \ref{hadronframe}),
where $p^\mu$ and $q^\mu$ are collinear and take the following form: 
\begin{align}
&p^\mu = \frac{Q}{2 x_{bj}}(1,0,0,1),\\
&q^\mu = (0,0,0,-Q). 
\end{align}
Defining the azimuthal angles for the hadron plane and the lepton plane
as $\chi$ and $\phi$, respectively, as shown in Fig. \ref{hadronframe}, 
$P_h^\mu$ and $\ell^\mu$ can be written as 
\beq
&& P_h^\mu = \frac{z_f Q}{2} \left(
1 + \frac{q_T^2}{Q^2}, \frac{2 q_T}{Q} \cos{\chi},
\frac{2 q_T}{Q} \sin{\chi}, -1 + \frac{q_T^2}{Q^2}
\right),
\label{Ph}\\
&&\ell^\mu = \frac{Q}{2}(\cosh{\psi}, \sinh{\psi}\cos{\phi},
\sinh{\psi}\sin{\phi},-1 ),
\label{lepmom}
\eeq
where $\psi$ is defined by
\beq
\cosh{\psi} \equiv \frac{2x_{bj} S_{ep}}{Q^2} -1.  
\label{coshpsi}
\eeq
With this parameterization, 
the transverse momentum of the hyperon $P_{hT}$
is given by
$P_{hT}= z_f q_T$. 

For the calculation of the cross section, 
we introduce four axes by
\begin{align}
\label{txyz}
 & T^\mu \equiv \frac{1}{Q}(q^\mu + 2 x_{bj} p^\mu)=(1,0,0,0),{\nonumber}\\
&Z^\mu \equiv - \frac{q^\mu}{Q} = (0,0,0,1), \nonumber\\
 &X^\mu \equiv \frac{1}{q_T} \left[
\frac{P_h^\mu}{z_f} - q^\mu - \left(
1+ \frac{q_T^2}{Q^2} 
\right)x_{bj} p^\mu
\right] = (0, \cos{\chi}, \sin{\chi}, 0) , {\nonumber}\\
 & Y^\mu \equiv \epsilon^{\mu \nu \rho \sigma}Z_\nu T_\rho X_\sigma  
= (0, - \sin{\chi}, \cos{\chi},0),
\end{align}
where the actual form in the hadron frame is given after the last equality in each equation. 
The final hyperon resides in the $XZ$-plane and the transverse spin vector of the hyperon
can be written as
\begin{align}
\label{spinvector}
 S_\perp^\mu = \cos{\theta}\cos{\Phi_S}X^\mu + \sin{\Phi_S}Y^\mu - \sin{\theta}\cos{\Phi_S}Z^\mu, 
\end{align}
where $\theta$ is the polar angle of $\vec{P_h}$ as measured from the $Z$-axis
and $\Phi_S$ is the azimuthal angle of $\vec{S}_\perp$ around $\vec{P}_h$
as measured from the $XZ$-plane.   
From (\ref{Ph}), the polar angle $\theta$ is written as
\begin{align}
 \cos{\theta} &= \frac{P_{hz}}{|\vec P_h|}
= \frac{q_T^2 - Q^2}{q_T^2 + Q^2}, \\
 \sin{\theta}&=   \frac{P_{h T}}{|\vec P_h |}
= \frac{2 q_T Q}{q_T^2 + Q^2}. 
\end{align}

With the kinematical variables defined above,
the polarized differential cross section for (\ref{epeLX}), $\sigma\equiv\sigma(p,\ell,\ell',P_h,S_\perp)$, 
takes the following form:
\begin{align}
\label{Xsec}
 \frac{\drm^6 \sigma}{\drm x_{bj} \drm Q^2 \drm z_f \drm q_T^2 \drm \phi \drm \chi}
= \frac{\alpha_{em}^2}{128 \pi^4 x_{bj}^2 S_{ep}^2 Q^2}z_f
L^{\rho\sigma}(\ell , \ell')W_{\rho\sigma}(p,q,P_h) ,
\end{align}
where $\alpha_{em} = e^2/(4\pi)$ is the QED coupling constant, 
$ L^{\rho\sigma}=2(\ell^\rho \ell'^\sigma + \ell^\sigma \ell'^\rho)-Q^2 g^{\rho\sigma}$
is the leptonic tensor and $W_{\rho\sigma}$ is the hadronic tensor. 
Although there are two azimuthal angles, $\phi$ and $\chi$, the cross section depends on 
the relative angle $\varphi \equiv \phi - \chi$ only.
Therefore it can be expressed in terms of 
$S_{ep}$, $Q^2$, $x_{bj}$, $z_f$, $q_T^2$, $\varphi$ and $\Phi_S$.
\subsection{Hadronic tensor}
\begin{figure}[ht]
 \begin{center}
\includegraphics[scale=0.93]{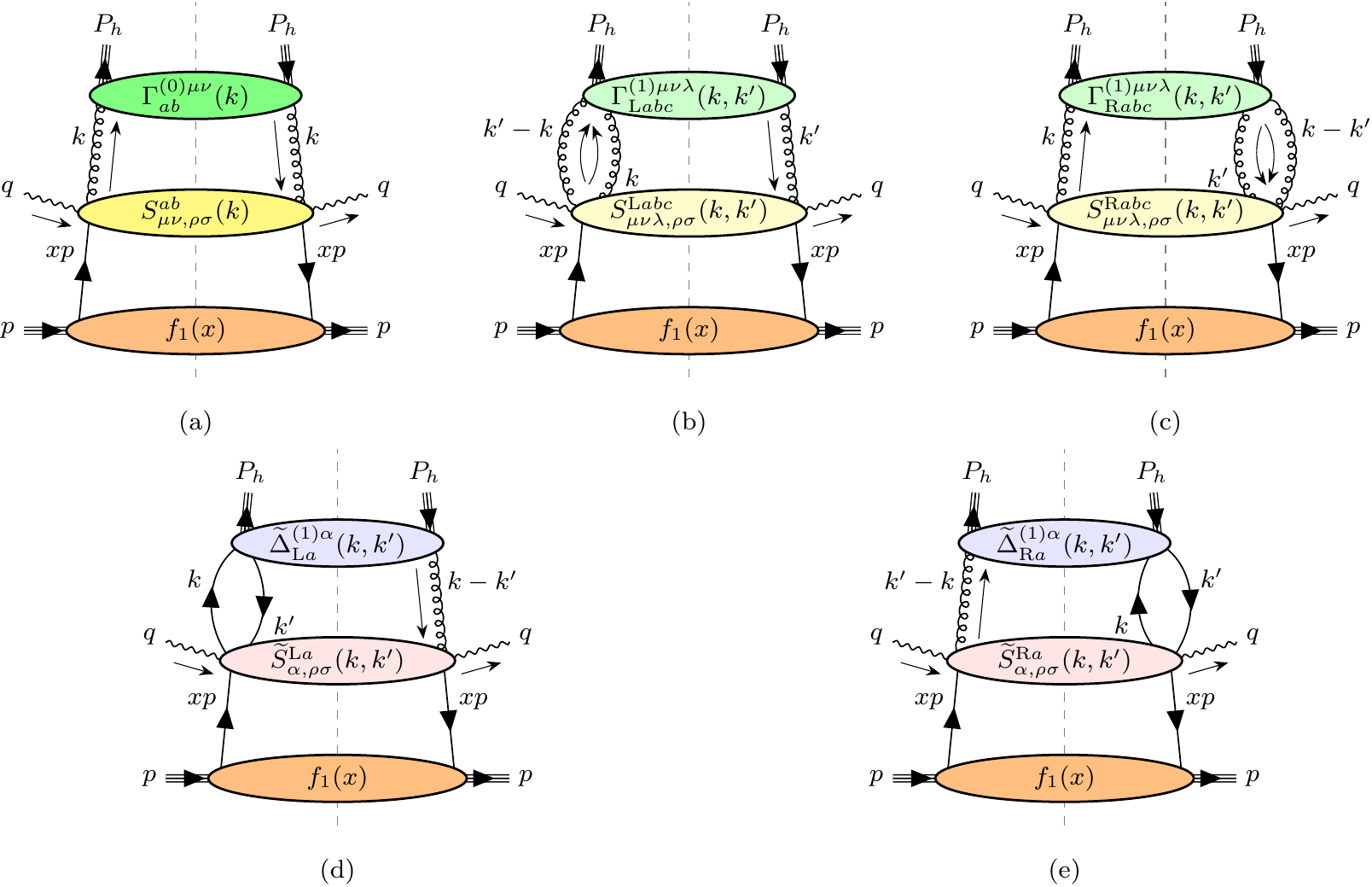}
 \end{center}
 \caption{Cut diagrams for the twist-3 gluon fragmentation contribution to $ep\to e\Lambda^\uparrow X$.  
In each diagram, the lower blob represents the unpolarized quark distribution, the middle one 
represents the partonic hard cross section and the upper one represents the fragmentation matrix elements 
for the final hyperon.}
\label{genericdiagrams}
 \end{figure}

We now calculate the twist-3 gluon FF contribution to (\ref{Xsec})
following the formalism developed for $pp \to \Lambda^\uparrow X$\cite{Koike:2021awj}.  
It occurs as a nonpole contribution from the hard part
as in the case of other twist-3 fragmentation contributions in 
$ep^{\uparrow} \to e \pi X$ \cite{Kanazawa:2013uia} and
$pp \to \Lambda^{\uparrow} X$ \cite{Koike:2017fxr, Koike:2021awj}. 
We first factorize the twist-2 unpolarized quark DFs $f_1(x)$ from the hadronic tensor 
$W_{\rho\sigma}(p,q,P_h)$: 
\begin{align}
\label{factor}
W_{\rho\sigma}(p,q,P_h) = \int \frac{\drm x}{x}f_1(x) w_{\rho\sigma}(xp,q,P_h),
\end{align}
where $x$ is the momentum fraction of the quark in the proton,
and we have omitted the factor associated with the quark's fractional electric charge as well as
summation over quark flavors.  
Up to twist-3, 
$w_{\rho\sigma}$ receives contribution from 
the 2-gluon, 3-gluon and quark-antiquark-gluon correlation 
functions corresponding to (a)-(e) of Fig. \ref{genericdiagrams}:  
\begin{align}
  w_{\rho \sigma} \equiv w_{\rho \sigma}^{\rm (a)} + w_{\rho \sigma}^{\rm (b)} + w_{\rho \sigma}^{\rm (c)} 
+ w_{\rho \sigma}^{\rm (d)} + w_{\rho \sigma}^{\rm (e)}, 
\end{align} 
where each term can be written as 
\begin{align}
w_{\rho \sigma}^{\rm (a)} &= \int \frac{\drm^4 k}{(2\pi)^4}\Gamma^{(0)\mu\nu}_{ab}(k) 
S^{ab}_{\mu \nu , \rho \sigma}(k), \\[7pt]
w_{\rho \sigma}^{\rm (b)}&=   \frac{1}{2}\iint \frac{\drm^4 k}{(2\pi)^4}\frac{\drm^4 k'}{(2\pi)^4} 
\Gamma^{(1)\mu \nu \lambda}_{\Lrm abc}(k,k')
S^{\Lrm abc}_{\mu \nu \lambda, \rho \sigma }(k,k'), 
\label{gggL}\\[7pt]
 w_{\rho \sigma}^{\rm (c)}&=   \frac{1}{2}\iint \frac{\drm^4 k}{(2\pi)^4}\frac{\drm^4 k'}{(2\pi)^4} 
\Gamma^{(1)\mu \nu \lambda}_{\Rrm abc}(k,k')
S^{\Rrm abc}_{\mu \nu \lambda, \rho \sigma }(k,k'), 
\label{gggR}\\[7pt]
w_{\rho \sigma}^{\rm (d)}&=  \Tr{} \iint \frac{\drm^4 k}{(2\pi)^4}\frac{\drm^4 k'}{(2\pi)^4}
\widetilde{\Delta}^{(1)\alpha}_{\Lrm a}(k,k')\widetilde{S}^{\Lrm a}_{\alpha, \rho \sigma}(k,k'), 
\label{tildeL}\\[7pt]
 w_{\rho \sigma}^{\rm (e)}&=  \Tr{} \iint \frac{\drm^4 k}{(2\pi)^4}\frac{\drm^4 k'}{(2\pi)^4}
\widetilde{\Delta}^{(1)\alpha}_{\Rrm a}(k,k')\widetilde{S}^{\Rrm a}_{\alpha, \rho \sigma}(k,k').  
\label{tildeR}
\end{align}
Here 
$S^{ab}_{\mu\nu, \rho\sigma}(k)$, 
$S^{\Lrm(\Rrm) abc}_{\mu\nu\lambda, \rho\sigma}(k,k')$, 
and
$\widetilde{S}^{\Lrm(\Rrm) a}_{\alpha,\rho\sigma}(k,k')$
represent the partonic hard parts
with
$k$ and $k'$ the momenta of partons fragmenting into the final hyperon, 
and the dependence on $q$ is suppressed for simplicity.
$\Gamma^{(0)\mu\nu}_{ab}$,
$\Gamma^{(1)\mu\nu\lambda}_{\Lrm(\Rrm) abc}$
and
$\widetilde{\Delta}^{(1)\alpha}_{\Lrm(\Rrm) a}$ 
denote the fragmentation matrix elements defined as
\begin{align}
 \Gamma^{(0)\mu\nu}_{ab}(k) &=
\sum_{X}\int\drm^4 \xi \Exp{- \irm k \cdot \xi}
\bra{0}A^{\nu}_{b}(0)\ket{hX}\bra{hX}A^{\mu}_{a}(\xi)\ket{0}, \\[7pt]
 \Gamma^{(1)\mu\nu\lambda}_{\Lrm abc}(k,k')&= 
\sum_{X}\iint\drm^4 \xi \drm^4 \eta \Exp{- \irm k \cdot \xi}\Exp{-\irm(k'-k)\cdot \eta}
\bra{0}A^{\nu}_{b}(0)\ket{hX}\bra{hX}A^{\mu}_{a}(\xi)gA^{\lambda}_c(\eta)\ket{0}, \\[7pt]
 \Gamma^{(1)\mu\nu\lambda}_{\Rrm abc}(k,k')&= 
\sum_{X}\iint\drm^4 \xi \drm^4 \eta \Exp{- \irm k \cdot \xi}\Exp{-\irm(k'-k)\cdot \eta}
\bra{0}A^{\nu}_{b}(0)gA^{\lambda}_c(\eta)\ket{hX}\bra{hX}A^{\mu}_{a}(\xi)\ket{0}, 
\label{qqbargL}\\[7pt]
 \widetilde{\Delta}^{(1)\alpha}_{\Lrm a,ij}(k,k')&= 
\sum_{X}\iint\drm^4 \xi \drm^4 \eta \Exp{- \irm k \cdot \xi}\Exp{-\irm(k'-k)\cdot \eta}
\bra{0}gA^{\alpha}_{a}(\eta)\ket{hX}\bra{hX}\psi_i(0)\bar \psi_j(\xi)\ket{0}, 
\label{qqbargR}\\[7pt] 
 \widetilde{\Delta}^{(1)\alpha}_{\Rrm a,ij}(k,k')&= 
\sum_{X}\iint\drm^4 \xi \drm^4 \eta \Exp{- \irm k \cdot \xi}\Exp{-\irm(k'-k)\cdot \eta}
  \bra{0}\psi_i(0)\bar \psi_j(\xi)\ket{hX}\bra{hX}gA^{\alpha}_{a}(\eta)\ket{0}.
\end{align}
The contribution with two parton lines in the left (right) of the cut
in Fig. \ref{genericdiagrams} (b)-(e) are characterized
by the symbol $\Lrm$ ($\Rrm$) in the hard parts and the fragmentation matrix elements.  
The superscripts $(0)$ and $(1)$ indicate the order of the gauge coupling $g$
corresponding, respectively, to the 2-parton 
and 3-parton correlation functions.  
The factor $1/2$ in (\ref{gggL}) and (\ref{gggR}) takes into account the exchange symmetry in the 
corresponding matrix element.  
In (\ref{tildeL}) and (\ref{tildeR}), the hard parts and the fragmentation matrix elements
are matrices both in color and spinor spaces for the quark and ${\rm Tr}$ indicates trace over 
both indices. 
The hard parts and the fragmentation matrix elements satisfy 
$ \Gamma^{(1)\mu \nu \lambda}_{\Rrm abc}(k,k')= 
\Gamma^{(1)\nu \mu \lambda}_{\Lrm bac} (k',k)^*$ ,
$ \widetilde \Delta^{(1)\alpha}_{\Rrm a}(k,k')=  
\gamma^0 \widetilde \Delta^{(1)\alpha}_{\Lrm a}(k',k)^\dagger\gamma^0$ ,
$S^{\Rrm abc}_{ \mu \nu \lambda,\rho\sigma}(k,k')= S^{\Lrm bac}_{ \nu\mu\lambda,\sigma\rho}(k',k)^*$
and
$\widetilde S^{\Rrm a}_{\alpha,\rho\sigma}(k,k')=  \gamma^0 \widetilde S^{\Lrm a}_{\alpha,\sigma\rho}(k',k)^\dagger \gamma^0$.
We thus have 
\beq
w_{\rho\sigma} = w^{\rm (a)}_{\rho\sigma} + 2 \Re\, w^{\rm (b)}_{\rho\sigma} 
+ 2 \Re\, w^{\rm (d)}_{\rho\sigma}.
\eeq

\begin{figure}[th]
 \begin{center}
\includegraphics[scale=0.95]{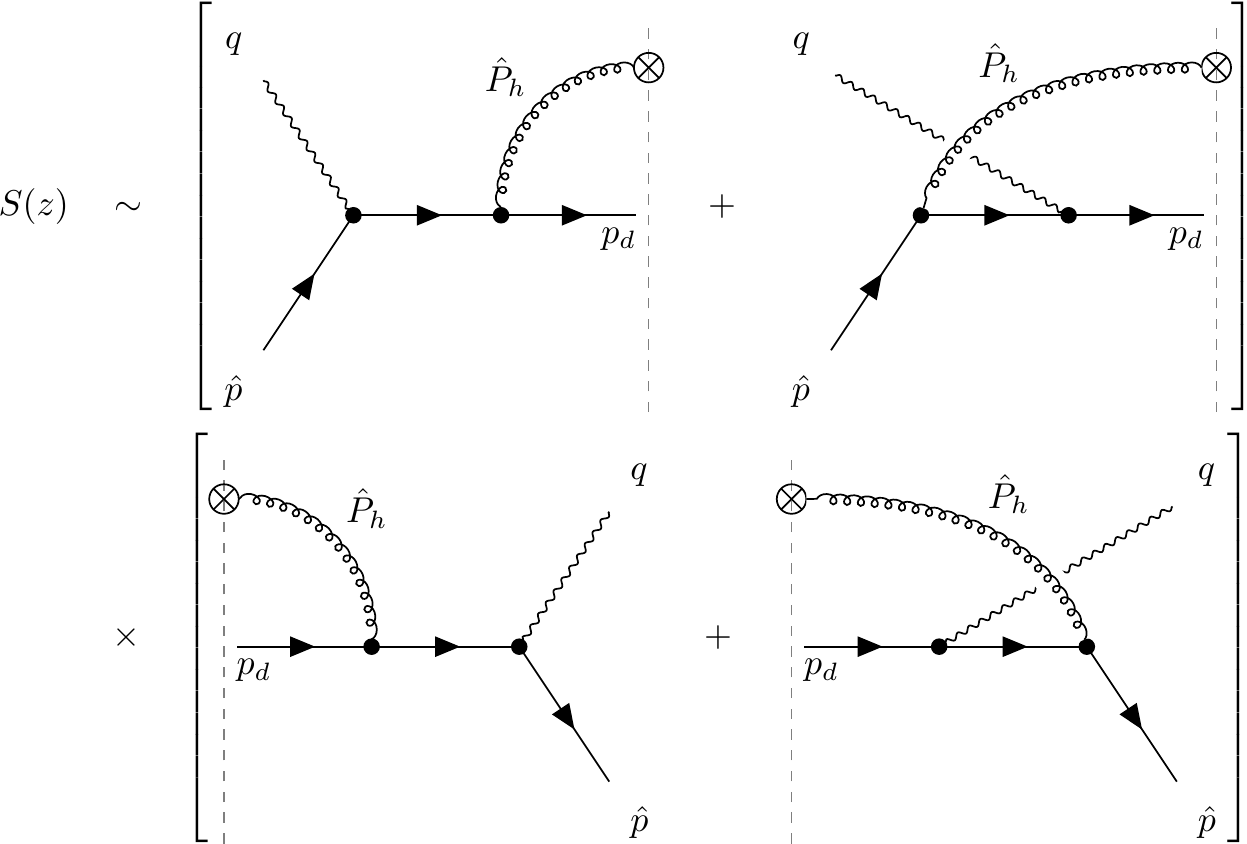}
 \end{center}
 \caption{The lowest order
Feynman diagrams for $S_{\alpha\beta,\rho\sigma}\left({1\over z}\right)$ in (\ref{hadronic tensor}).
We set $\hat p \equiv xp$ and $ \hat P_h \equiv P_h / z$.
The symbol $\otimes$ indicates the fragmentation to the final hadron 
and $p_d$ is the momentum of an unobserved parton in the final state.
}
\label{hardS}
\end{figure}

\begin{figure}[th]
 \begin{center}
\includegraphics[scale=0.95]{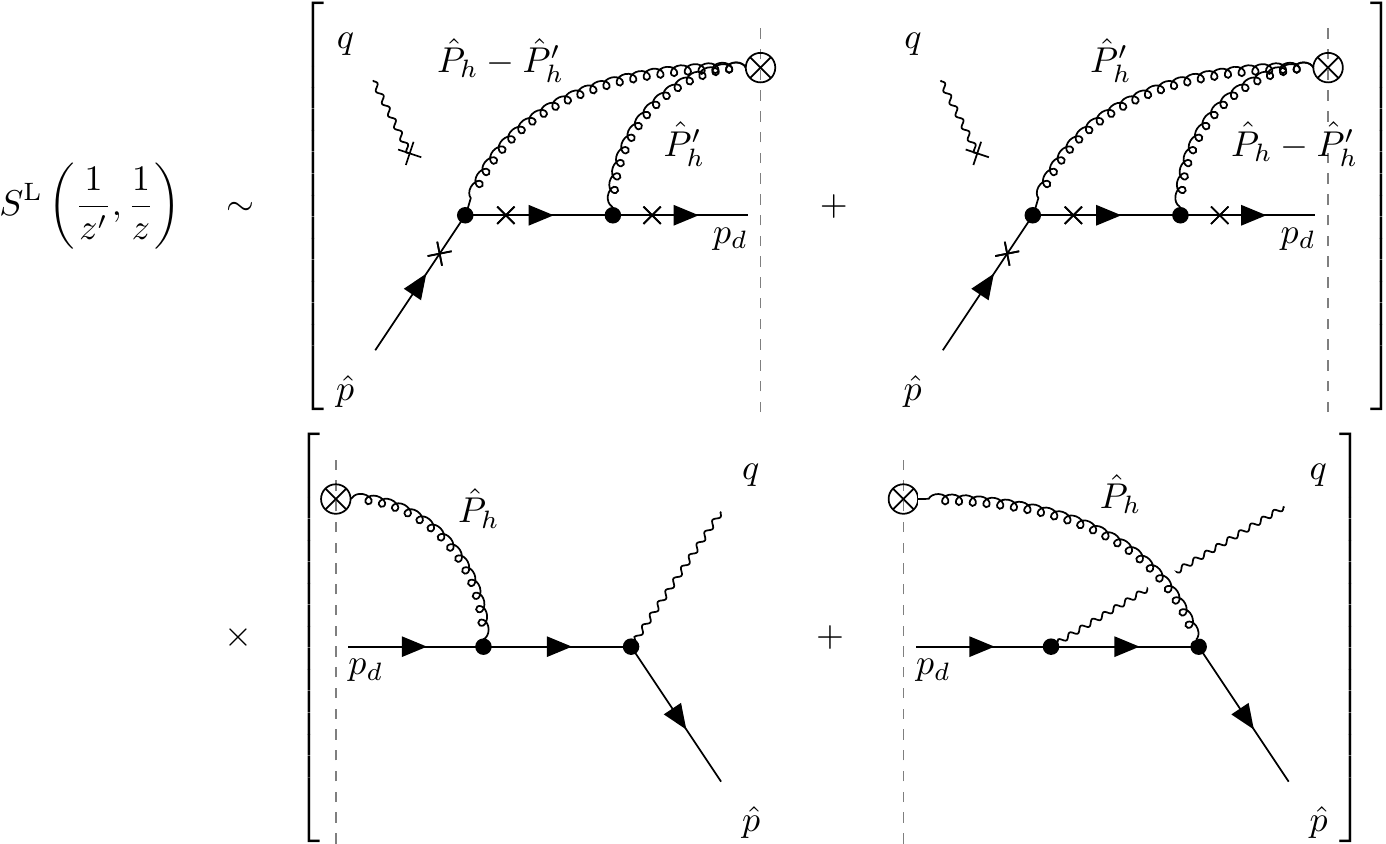}
 \end{center}
 \caption{
 The lowest order Feynman diagrams for 
$S^{\Lrm abc}_{\alpha\beta\gamma,\rho\sigma}\left(\frac{1}{z'},\frac{1}{z}\right)$ in (\ref{hadronic tensor}). 
We set $\hat P'_h \equiv P_h / z'$.
Three crosses ($\times$) on the quark line in the upper diagrams
indicates that the virtual photon line with a cross at one end needs to be attached to one of these crosses,
and all three diagrams have to be included.  
Thus the number of diagrams 
for $S^{\Lrm abc}_{\alpha\beta\gamma,\rho\sigma}$
is $ (3 + 3) \times 2 = 12$.  The meaning of the other symbols are the same as 
that in Fig. \ref{hardS}.  
}
\label{hardSL}
\end{figure}

\begin{figure}[th]
 \begin{center}
\includegraphics[scale=0.95]{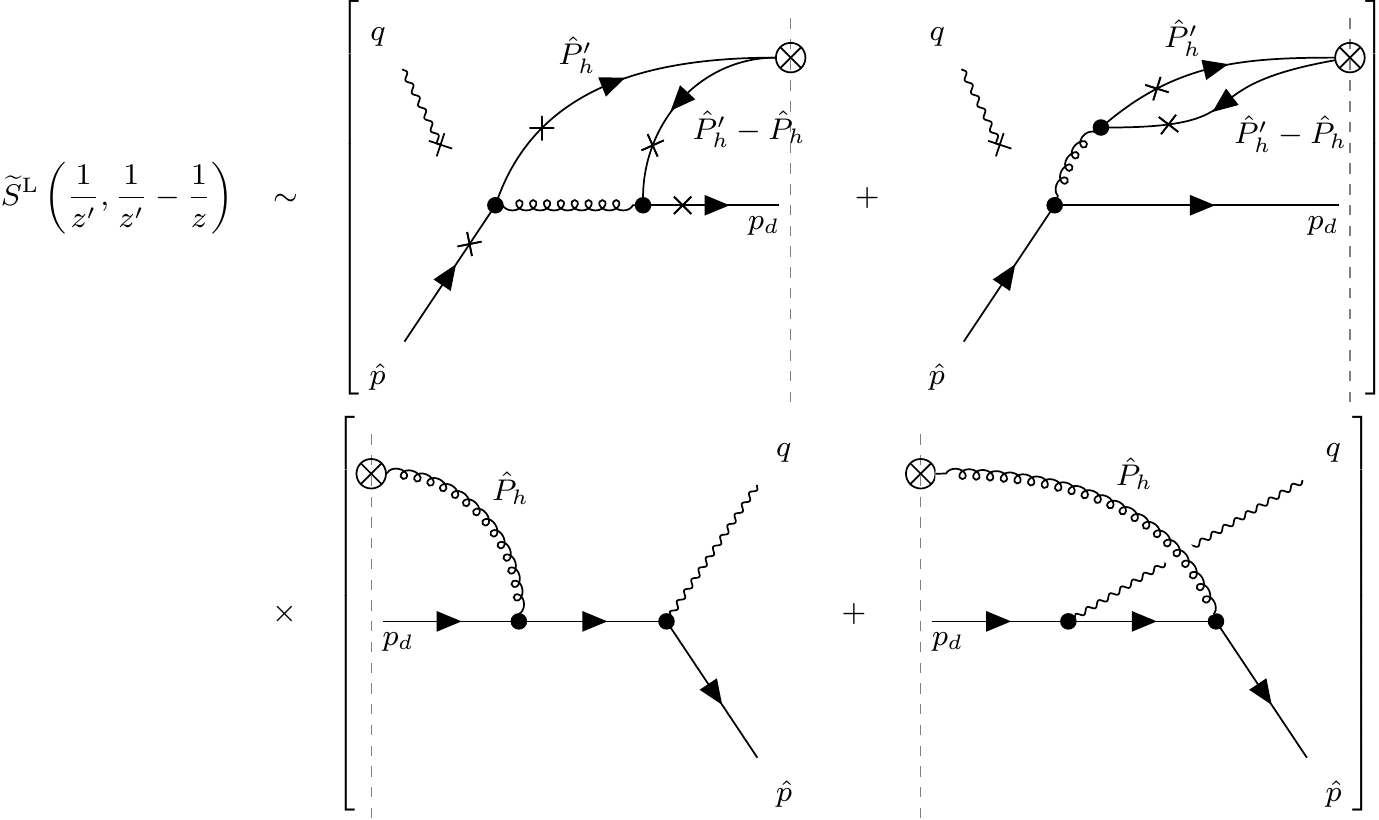}
 \end{center}
 \caption{The lowest order Feynman diagrams for
$\widetilde{S}^{\Lrm}_{\alpha,\rho\sigma}\left(\frac{1}{z'}, \frac{1}{z'}-\frac{1}{z}\right)$.
 The meaning of the symbols is the same as that in Fig. \ref{hardSL}.  
The total number of diagrams for $\widetilde{S}^{\Lrm}_{\alpha,\rho\sigma}$ is $(4 + 2) \times 2 = 12$.}
\label{hardSLtilde}
\end{figure}

To extract the twist-3 contribution to $ep\to e\Lambda^\uparrow X$
we apply the collinear expansion to the hard part,
$S^{ab}_{\mu\nu,\rho\sigma}$, $S^{\Lrm abc}_{\mu\nu\lambda,\rho\sigma}$ 
and $\widetilde{S}^{\Lrm a}_{\alpha,\rho\sigma}$, 
with respect to the momenta $k$ and $k'$
around $P_h/z$ and $P_h/z'$, respectively, taking into account of the following 
Ward-Takahashi (WT) identities\cite{Koike:2021awj}:
\begin{align}
& k^\mu S^{ab}_{\mu \nu,\rho\sigma}(k) = k^\nu S^{ab}_{\mu \nu,\rho\sigma}(k) = 0, 
\label{WTI1}\\[7pt]
& k^\mu S^{\Lrm abc}_{\mu\nu\lambda,\rho\sigma}(k,k')=\frac{\irm f^{abc}}{N^2 -1}S_{\lambda \nu,\rho\sigma}(k'), \\[7pt]
&k'^\nu S^{\Lrm abc}_{\mu \nu\lambda,\rho\sigma}(k,k')=  0, \\[7pt]
&(k'-k)^\lambda S^{\Lrm abc}_{\mu\nu\lambda,\rho\sigma}(k,k')=   \frac{- \irm f^{abc}}{N^2 -1}S_{\mu \nu,\rho\sigma}(k'), \\[7pt]
& (k-k')^\alpha \widetilde S^{\Lrm a}_{\alpha,\rho\sigma}(k,k') =0,
\label{WTI2}
\end{align}
where $S_{\mu\nu,\rho\sigma}(k)\equiv \delta^{ab}S^{ab}_{\mu\nu,\rho\sigma}(k)$.
We note that unlike the case for $pp\to\Lambda^\uparrow X$ no ghost-like terms appear
in the WT identities (\ref{WTI1})-(\ref{WTI2}) for the present case.  
This way, one obtains the hadronic tensor $w_{\rho\sigma}$
in terms of the gauge invariant FFs as (See eq. (51) of
\cite{Koike:2021awj} and eq. (56) of \cite{Kanazawa:2013uia})
\beq
&&w_{\rho\sigma}=
\Omega^{\alpha}_{\;\; \mu} \Omega^{\beta}_{\;\; \nu}\int \drm \left( \frac{1}{z} \right) z^2
{\widehat \Gamma^{\mu\nu}(z)}
{S_{\alpha\beta,\rho\sigma}\left({1\over z}\right)}
 {\nonumber}\\[7pt]
&&\qquad - \irm \Omega^{\alpha}_{\;\; \mu}\Omega^{\beta}_{\;\; \nu}\Omega^{\gamma}_{\;\; \lambda}
\int\drm \left(\frac{1}{z}\right)z^2 
{\widehat \Gamma^{\mu\nu\lambda}_{\partial}(z) }
\left. \frac{\partial 
{S_{\alpha\beta , \rho\sigma}(k)}
}{\partial k^\gamma}\right|_{k=P_h/z} {\nonumber}\\[7pt]
&&\qquad + \Re \left[
i \Omega^{\alpha}_{\;\; \mu}\Omega^{\beta}_{\;\; \nu}\Omega^{\gamma}_{\;\; \lambda}
\iint \drm \left(\frac{1}{z}\right)\drm \left(\frac{1}{z'}\right)zz' \frac{1}{1/z-1/z'} \right.
{\nonumber}\\[4pt]
&& \qquad\qquad\times \left. 
\left\{ -\frac{i f^{abc}}{N}
{\widehat \Gamma^{\mu\nu\lambda}_{FA}\left(\frac{1}{z'},\frac{1}{z}\right) }
+\frac{N d^{abc}}{N^2-4}
{\widehat \Gamma^{\mu\nu\lambda}_{FS}\left(\frac{1}{z'},\frac{1}{z}\right)}
\right\}
{S^{\Lrm abc}_{\alpha\beta\gamma,\rho\sigma}\left(\frac{1}{z'},\frac{1}{z}\right)}
\right]{\nonumber}\\[7pt]
&& \qquad
+ \Re\left[ i \Omega^{\alpha}_{\;\; \mu} \iint \drm\left(\frac{1}{z}\right) 
\drm\left(\frac{1}{z'}\right) z\,
{\rm Tr}_{\rm s}\left\{ 
{\widetilde \Delta^{\mu}\left(\frac{1}{z'},\frac{1}{z'}-\frac{1}{z}\right)}
{\widetilde S^{\Lrm}_{\alpha,\rho\sigma}\left(\frac{1}{z'},\frac{1}{z'}-\frac{1}{z}\right)}
\right\}\right],
\label{hadronic tensor}
\eeq
where 
$\widehat \Gamma^{\mu\nu}(z)$,
$\widehat \Gamma^{\mu\nu\lambda}_\partial (z)$,
$\widehat \Gamma^{\mu\nu\lambda}_{FA}\left(\frac{1}{z'},\frac{1}{z}\right)$,
$\widehat \Gamma^{\mu\nu\lambda}_{FS}\left(\frac{1}{z'},\frac{1}{z}\right)$
and
$\widetilde \Delta^{\mu}\left(\frac{1}{z'},\frac{1}{z'}-\frac{1}{z}\right)$
are given by
(\ref{intrinsic}), (\ref{kinematical}), (\ref{dynamicalFA}), (\ref{dynamicalFS}) and 
(\ref{dynamicalDelta}).
For the hard part we have used the notation 
$S_{\alpha\beta,\rho\sigma}\left({1\over z}\right)$ for 
$S_{\alpha\beta,\rho\sigma}\left({P_h\over z}\right)$ and
$S^{\Lrm abc}_{\alpha\beta\gamma,\rho\sigma}\left(\frac{1}{z'},\frac{1}{z}\right)$ for 
$S^{\Lrm abc}_{\alpha\beta\gamma,\rho\sigma}\left(\frac{P_h}{z'},\frac{P_h}{z}\right)$, 
etc, suppressing $P_h$ for short.  
In the last term of (\ref{hadronic tensor}), 
$\widetilde{S}^{\Lrm}_{\alpha,\rho\sigma}$ is defined from 
$\widetilde{S}^{\Lrm a}_{\alpha, \rho \sigma}$ in (\ref{tildeL}) by
$\displaystyle \left(\widetilde{S}^{\Lrm a}_{\alpha, \rho \sigma} \right)_{rs}
= {1\over 2N}t^a_{rs}\widetilde{S}^{\Lrm}_{\alpha,\rho\sigma}$,
where $r,s$ indicates the color indices for the quark, and 
${\rm Tr}_{\rm s}$ denotes the trace in the spinor space. 
The LO diagrams for the hard parts of Figs. \ref{genericdiagrams}(a), (b) and (d)
are, respectively, shown in Figs. \ref{hardS}, \ref{hardSL} and \ref{hardSLtilde}.  
It is easy to show that the hadronic tensor $w_{\rho \sigma}$ satisfies 
the electromagnetic gauge invariance,
$q^\rho w_{\rho\sigma} = q^\sigma w_{\rho\sigma} = 0$,
owing to the WT identity in QED.


\subsection{Spin dependent cross section}
The calculation of
$L^{\rho\sigma}W_{\rho\sigma}$ in ({\ref{Xsec}) can be done in the same way 
as \cite{Koike:2022ddx}:  
$W^{\rho\sigma}$ can be expanded in terms of the six tensors\cite{Meng:1991da}
$\mathscr{V}_k^{\rho\sigma}$ ($k=1,\cdots,4, 8 ,9)$ defined by
\begin{align}
  &\mathscr{V}^{\mu \nu}_1 = X^\mu X^\nu + Y^\mu Y^\nu ,&
& \mathscr{V}^{\mu \nu}_2 = g^{\mu \nu}+ Z^\mu Z^\nu, &
& \mathscr{V}^{\mu \nu}_3 = T^\mu X^\nu + X^\mu T^\nu ,
{\nonumber}\\
  &\mathscr{V}^{\mu \nu}_4 = X^\mu X^\nu - Y^\mu Y^\nu ,&
& \mathscr{V}^{\mu \nu}_8 = T^\mu Y^\nu  + Y^\mu T^\nu,&
 &\mathscr{V}^{\mu \nu}_9 = X^\mu Y^\nu + Y^\mu X^\nu.  
\end{align}
By introducing the inverses of $\mathscr{V}_k^{\rho\sigma}$, $\tilde{\mathscr{V}}_k^{\rho\sigma}$
satisfying $\mathscr{V}_k^{\rho\sigma}\tilde{\mathscr{V}}_{k'\,\rho\sigma}=\delta_{kk'}$, 
as
\begin{align}
& \tilde{\mathscr{V}}^{\mu \nu}_1 = \frac{1}{2}(2T^\mu T^\nu + X^\mu X^\nu + Y^\mu Y^\nu) ,\quad
\tilde{\mathscr{V}}^{\mu \nu}_2 = T^\mu T^\nu,  &
&\tilde{\mathscr{V}}^{\mu \nu}_3 = - \frac{1}{2}(T^\mu X^\nu + X^\mu T^\nu), {\nonumber}\\
&\tilde{\mathscr{V}}^{\mu \nu}_4 = \frac{1}{2}(X^\mu X^\nu - Y^\mu Y^\nu) , \quad
  \tilde{\mathscr{V}}^{\mu \nu}_8 = -\frac{1}{2}(T^\mu Y^\nu  + Y^\mu T^\nu), &
&\tilde{\mathscr{V}}^{\mu \nu}_9 = \frac{1}{2}(X^\mu Y^\nu + Y^\mu X^\nu), 
\end{align}
$W^{\mu\nu}$ can be expanded as
\begin{align}
W^{\mu \nu} = \sum_{k=1,\cdots,4,8,9}
\mathscr{V}^{\mu \nu}_k[W_{\rho \sigma}\tilde{\mathscr{V}}^{\rho \sigma}_k].  
\end{align}
Then one obtains 
\beq
 L^{\mu \nu}W_{\mu \nu}=
\sum_{k=1,\cdots 4,8,9}
[L_{\mu \nu}\mathscr{V}^{\mu \nu}_k]
[W_{\rho \sigma}\tilde{\mathscr{V}}^{\rho \sigma}_k]
= Q^2 \sum_{k=1, \cdots 4,8,9}
\mathscr{A}_k(\phi-\chi)
[W_{\rho \sigma}\tilde{\mathscr{V}}^{\rho \sigma}_k],
\label{LWcont}
\eeq
where
$\mathscr{A}_k(\varphi) \equiv L_{\mu \nu}\mathscr{V}^{\mu\nu}_k/Q^2$
are given by
\begin{align}
& \mathscr{A}_1(\varphi) = 1 + \cosh^2{\psi}, &
&\mathscr{A}_2(\varphi) = -2, &
&\mathscr{A}_3(\varphi) = - \cos{\varphi}\sinh{2\psi},{\nonumber} \\
 &\mathscr{A}_4(\varphi) = \cos{2\varphi} \sinh^2{\psi} ,&
&\mathscr{A}_8(\varphi)  = - \sin{\varphi}\sinh{2\psi}, &
&\mathscr{A}_9(\varphi) = \sin{2\varphi} \sinh^2{\psi},
\label{Aks}
\end{align}
with $\psi$ defined in (\ref{coshpsi}).  
From (\ref{LWcont}) and (\ref{Aks}), one sees the cross section
can be decomposed into the five structure functions with different azimuthal dependences
which are carried by the $\mathscr{A}_k(\varphi)$s.  
Substituting (\ref{intrinsic}), (\ref{kinematical}), (\ref{dynamicalFA}), 
(\ref{dynamicalFS}) and (\ref{dynamicalDelta})
into (\ref{hadronic tensor}), we find that the cross section (\ref{Xsec}) takes the
following structure:
\begin{align}
&\frac{\drm^6 \sigma}{\drm x_{bj}\drm Q^2 \drm z_f \drm q_\Trm^2
\drm \phi \drm \chi}  {\nonumber}\\[7pt]
&= \frac{\alpha_{em}^2 \alpha_s M_h}{16\pi^2 x_{bj}^2 S_{ep}^2Q^2}
\sum_{k}
 \mathscr{A}_k(\varphi) \mathcal{S}_k
\iint \drm x \drm \left(\frac{1}{z}\right) \frac{z^3}{x} f_1(x)
\delta \left(\frac{q_T^2}{Q^2} - \left(1-\frac{1}{\hat x} \right)\left(1-\frac{1}{\hat z} \right) \right) 
 {\nonumber}\\[7pt]
&\times\left\{
\frac{1}{z}\Delta\widehat G_{3\bar T}(z)\hat \sigma^k_{int}
+ \widehat G^{(1)}_T(z)\hat \sigma^k_{NDG}
+\frac{1}{z}\frac{\partial \widehat G_{T}^{(1)}(z)}{\partial (1/z)}\hat \sigma^k_{DG}
+\Delta \widehat H^{(1)}_T (z)\hat \sigma^k_{NDH}
+\frac{1}{z}\frac{\partial \Delta\widehat H_{T}^{(1)}(z)}{\partial (1/z)}\hat \sigma^k_{DH}
\right.
{\nonumber}\\[7pt]
&+ \frac{1}{2}\int \drm\left(\frac{1}{z'}\right)
\left[
\sum_{i=1}^{3} \Im \widehat N_i\left(\frac{1}{z'},\frac{1}{z}\right)
\left(
\frac{1}{1/z-1/z'}\hat \sigma^{-(1)}_{i,k}
+ \frac{1}{z}\left(\frac{1}{1/z-1/z'}\right)^2 \hat\sigma^{-(2)}_{i,k}
+ z' \hat \sigma^{-(3)}_{i,k}
+ \frac{z'^2}{z}\hat\sigma^{-(4)}_{i,k}
\right)
\right.
{\nonumber}\\[7pt]
&\left.
+ \sum_{i=1}^{3} \widehat O_i\left(\frac{1}{z'},\frac{1}{z}\right)
\left(
\frac{1}{1/z-1/z'}\hat \sigma^{+(1)}_{i,k}
+ \frac{1}{z}\left(\frac{1}{1/z-1/z'}\right)^2 \hat\sigma^{+(2)}_{i,k}
+ z' \hat \sigma^{+(3)}_{i,k}
+ \frac{z'^2}{z}\hat\sigma^{+(4)}_{i,k}
\right)
\right]
{\nonumber}\\[7pt]
&
+ \int\drm\left(\frac{1}{z'}\right)\frac{2}{C_F} \left[ 
\Im 
{\widetilde D_{FT}\left(\frac{1}{z'},\frac{1}{z'}-\frac{1}{z} \right)} \right.
\left( 
{\hat \sigma^{k}_{DF}}
+ \frac{1}{z}\frac{1}{1/z-1/z'}
{\hat \sigma^k_{DF2}}
+ \frac{z'}{z}
{\hat \sigma^k_{DF3}}  \right.  {\nonumber}\\[7pt]
& \left. \qquad\qquad + \frac{1}{1-(1-q_T^2 / Q^2)z_f/z'}
{\hat \sigma^k_{DF4}}
+ \frac{1}{1-(1-q_T^2/Q^2)z_f(1/z-1/z')}
{\hat \sigma^k_{DF5}}
 \right) 
{\nonumber} \\[7pt]
&\qquad\qquad+\Im 
{\widetilde G_{FT}\left(\frac{1}{z'},\frac{1}{z'}-\frac{1}{z} \right)} 
\left( 
{\hat \sigma^{k}_{GF}}
+ \frac{1}{z}\frac{1}{1/z-1/z'}
{\hat \sigma^k_{GF2}}
+ \frac{z'}{z}
{\hat \sigma^k_{GF3}} \right.  {\nonumber}\\[7pt]
&\left. \qquad\qquad  \left. \left. + \frac{1}{1-(1-q_T^2 / Q^2)z_f/z'}
{\hat \sigma^k_{GF4}}
+ \frac{1}{1-(1-q_T^2/Q^2)z_f(1/z-1/z')}
{\hat \sigma^k_{GF5}}
 \right)
 \right]
\right\},
\label{Xsec2}
\end{align}
where 
\beq
\mathcal{S}_{1,2,3,4}\equiv\sin{\Phi_S}, \quad\mathcal{S}_{8,9}\equiv\cos{\Phi_S},\quad
\hat x = x_{bj}/x,\quad 
\hat z= z_f/z, 
\eeq
and we have set
\beq
\widehat N_3\left(\frac{1}{z'},\frac{1}{z}\right)
\equiv -\widehat N_2\left(\frac{1}{z}-\frac{1}{z'},\frac{1}{z}\right),\qquad
\widehat O_3\left(\frac{1}{z'},\frac{1}{z}\right)
\equiv \widehat O_2\left(\frac{1}{z}-\frac{1}{z'},\frac{1}{z}\right)
\label{N3O3}
\eeq
for convenience.   
Partonic hard parts for each FF can be computed from the
corresponding diagrams, Figs. \ref{hardS}, \ref{hardSL} and \ref{hardSLtilde}.  
We have reached 
the form (\ref{Xsec2}) based on the observation that the $z'$-dependence of the hard parts
for the dynamical FFs appears in the cross section only through the factors explicitly shown in
(\ref{Xsec2}) (See Appendix C of
\cite{Koike:2021awj}), and hence we can define
all the partonic hard cross sections $\hat{\sigma}$'s in (\ref{Xsec2}) as the functions
of $\hat{x}$, $\hat{z}$, $Q$ and $q_T$.   In addition we found by
explicit calculation of the LO diagrams that
\beq
&&\hat \sigma^{\pm(3)}_{i,k}= \hat \sigma^{\pm (1)}_{i,k},
\label{sigma1}\\
&&\hat \sigma^k_{DF}= \hat \sigma^k_{GF} = 0.
\label{sigma2}
\eeq

In order to transform the cross section (\ref{Xsec2}) into a more concise form, we note
the following points:
(I) Owing to the symmetry property under $1/z'\leftrightarrow 1/z-1/z'$ 
of $\widehat N_1$ and $\widehat O_1$ (\ref{symmetry})
and the relations (\ref{N3O3}), 
the terms of $\hat \sigma^{\pm(3)}_{i,k}$ and $\hat \sigma^{\pm(4)}_{i,k}$
can be combined, respectively, with those of $\hat \sigma^{\pm(1)}_{i,k}$
and $\hat \sigma^{\pm(2)}_{i,k}$, taking into account the relation (\ref{sigma1}).
(II) Using (\ref{eom1}), (\ref{rel1}) and (\ref{rel2}), one can eliminate
the intrinsic FF and the derivative of the kinematical FFs in favor of the kinematical
and the dynamical FFs.  
This way we finally obtain the twist-3 gluon FF contribution to 
$ep\to e\Lambda^\uparrow X$ as\footnote{As was noted at the end of Sec. 2, the kinematical FFs
$\widehat G^{(1)}_T (z)$ and $\Delta \widehat H^{(1)}_T (z)$
can, in principle, be eliminated in terms of the dynamical FFs.
}
\begin{align}
\label{result}
 &\frac{\drm^6 \sigma}{\drm x_{bj}\drm Q^2 \drm z_f \drm q_\Trm^2
\drm \phi \drm \chi}  {\nonumber}\\[7pt]
&= \frac{\alpha_{em}^2 \alpha_s M_h}{16\pi^2 x_{bj}^2 S_{ep}^2Q^2}
\sum_{k}
 \mathscr{A}_k(\phi - \chi) \mathcal{S}_k
\int^1_{x_{min}} {dx\over x}\int^1_{z_{min}} {dz\over z} 
z^2 f_1(x)
\delta \left(\frac{q_T^2}{Q^2} - \left(1-\frac{1}{\hat x} \right)\left(1-\frac{1}{\hat z} \right) \right) 
 {\nonumber}\\[7pt]
&\times\left\{
{\widehat G^{(1)}_T (z)}
{ \hat \sigma^k_G } 
+
 {\Delta \widehat H^{(1)}_T (z)}
{\hat \sigma^k_H} \right. {\nonumber}\\[7pt]
&+ \int\drm\left(\frac{1}{z'}\right)\left[
\frac{1}{1/z-1/z'}\Im \left(
{\widehat N_1 \left( \frac{1}{z'},\frac{1}{z} \right)}
{\hat\sigma^k_{N1}}
+
{\widehat N_2 \left( \frac{1}{z'},\frac{1}{z} \right)}
{\hat \sigma^k_{N2}}
+
{\widehat N_2 \left(\frac{1}{z}-\frac{1}{z'},\frac{1}{z} \right)}
{\hat \sigma^k_{N3}}
\right)
\right. {\nonumber}\\[7pt]
 &+ \frac{1}{z}\left(\frac{1}{1/z-1/z'}\right)^2 \Im\left(
{\widehat N_1\left( \frac{1}{z'},\frac{1}{z} \right)}
{\hat \sigma^k_{DN1}}
+
 {\widehat N_2 \left( \frac{1}{z'},\frac{1}{z} \right)}
{\hat \sigma^k_{DN2}}
+ 
{\widehat N_2 \left(\frac{1}{z}-\frac{1}{z'},\frac{1}{z} \right)}
{\hat \sigma^k_{DN3}}
\right) {\nonumber}\\[7pt]
&+ 
\frac{1}{1/z-1/z'}\Im \left(
{\widehat O_1 \left( \frac{1}{z'},\frac{1}{z} \right)}
{\hat\sigma^k_{O1}}
+
{\widehat O_2 \left( \frac{1}{z'},\frac{1}{z} \right)}
{\hat \sigma^k_{O2}}
+
{\widehat O_2 \left(\frac{1}{z}-\frac{1}{z'},\frac{1}{z} \right)}
{\hat \sigma^k_{O3}}
\right){\nonumber}\\[7pt]
&\left. + \frac{1}{z}\left(\frac{1}{1/z-1/z'}\right)^2 \Im\left(
{\widehat O_1\left( \frac{1}{z'},\frac{1}{z} \right)}
{\hat \sigma^k_{DO1}}
+
 {\widehat O_2 \left( \frac{1}{z'},\frac{1}{z} \right)}
{\hat \sigma^k_{DO2}}
+ 
{\widehat O_2 \left(\frac{1}{z}-\frac{1}{z'},\frac{1}{z} \right)}
{\hat \sigma^k_{DO3}}
\right)\right] {\nonumber}\\[7pt]
&
+ \int\drm\left(\frac{1}{z'}\right)\frac{2}{C_F} \left[ 
\Im 
{\widetilde D_{FT}\left(\frac{1}{z'},\frac{1}{z'}-\frac{1}{z} \right)} \right.
\left( 
{\hat \sigma^{k}_{DF1}}
+ \frac{1}{z}\frac{1}{1/z-1/z'}
{\hat \sigma^k_{DF2}}
+ \frac{z'}{z}
{\hat \sigma^k_{DF3}}  \right.  {\nonumber}\\[7pt]
& \left. + \frac{1}{1-(1-q_T^2 / Q^2)z_f/z'}
{\hat \sigma^k_{DF4}}
+ \frac{1}{1-(1-q_T^2/Q^2)z_f(1/z-1/z')}
{\hat \sigma^k_{DF5}}
 \right) 
{\nonumber} \\[7pt]
&+\Im 
{\widetilde G_{FT}\left(\frac{1}{z'},\frac{1}{z'}-\frac{1}{z} \right)} 
\left( 
\frac{1}{z}\frac{1}{1/z-1/z'}
{\hat \sigma^k_{GF2}}
+ \frac{z'}{z}
{\hat \sigma^k_{GF3}} \right.  {\nonumber}\\[7pt]
&\left.   \left. \left. + \frac{1}{1-(1-q_T^2 / Q^2)z_f/z'}
{\hat \sigma^k_{GF4}}
+ \frac{1}{1-(1-q_T^2/Q^2)z_f(1/z-1/z')}
{\hat \sigma^k_{GF5}}
 \right)
 \right]
\right\},
\end{align}
where the lower limits of $x$ and $z$ are, respectively, given by $x_{min} =x_{bj} 
\left(1 + \frac{z_f}{1-z_f}\frac{q_T^2}{Q^2}\right)$ and
$z_{min} = z_f \left( 1 + \frac{x_{bj}}{1-x_{bj}}\frac{q_T^2}{Q^2}\right)$.
The partonic hard cross sections which appear newly in (\ref{result})
are defined from those in (\ref{Xsec2}) as
\begin{align}
&\hat \sigma^k_G = \frac{1}{2} \hat \sigma^k_{int} + \hat \sigma^k_{NDG} + 2 \hat \sigma^k_{DG}, 
\label{Xsecparton1}\\
&\hat \sigma^k_H = \frac{1}{2} \hat \sigma^k_{int} + \hat \sigma^k_{NDH} + 4 \hat \sigma^k_{DH}, \\
&\hat \sigma^k_{N1} = 2 \hat \sigma^k_{int} + 4 \hat \sigma^k_{DG} + 8 \hat \sigma^k_{DH} , 
\label{sigmaN1}\\
&\hat \sigma^k_{N2} = \hat \sigma^k_{int} + 8 \hat \sigma^k_{DH} + \frac{1}{2}(\hat \sigma^{-(1)}_{2,k} -\hat \sigma^{-(1)}_{3,k}), \\
&\hat \sigma^k_{N3} = -\hat \sigma^k_{int} - 4\hat \sigma^k_{DG} + \frac{1}{2}(\hat \sigma^{-(1)}_{2,k} -\hat \sigma^{-(1)}_{3,k}) , \\
&\hat \sigma^k_{DN1} =  2\hat \sigma^k_{DG} +4\hat \sigma^k_{DH} + \frac{1}{2}(\hat \sigma^{-(2)}_{1,k} -\hat \sigma^{-(4)}_{1,k}), \\
&\hat \sigma^k_{DN2} = 2\hat \sigma^k_{DG} + 4\hat \sigma^k_{DH} + \frac{1}{2}(\hat \sigma^{-(2)}_{2,k} -\hat \sigma^{-(4)}_{3,k}), \\
&\hat \sigma^k_{DN3} = -4\hat \sigma^k_{DG} + \frac{1}{2}(\hat \sigma^{-(4)}_{2,k} -\hat \sigma^{-(2)}_{3,k}),\\
&\hat \sigma^k_{O1} =\hat \sigma^{+(1)}_{1,k} , \\
&\hat \sigma^k_{O2} =\frac{1}{2}(\hat \sigma^{+(1)}_{2,k} + \hat \sigma^{+(1)}_{3,k}),
\label{sigmaO2}\\
&\hat \sigma^k_{O3} = \frac{1}{2}(\hat \sigma^{+(1)}_{2,k} + \hat \sigma^{+(1)}_{3,k}) ,
\label{sigmaO3}\\
&\hat \sigma^k_{DO1} = \frac{1}{2}(\hat \sigma^{+(2)}_{1,k} +\hat \sigma^{+(4)}_{1,k}),\\
&\hat \sigma^k_{DO2} = \frac{1}{2}(\hat \sigma^{+(2)}_{2,k} + \hat \sigma^{+(4)}_{3,k}), \\
&\hat \sigma^k_{DO3} = \frac{1}{2}(\hat \sigma^{+(4)}_{2,k} +\hat \sigma^{+(2)}_{3,k}),\\
&\hat \sigma^k_{DF1} = -\hat \sigma^k_{int} - 
 2 \hat \sigma^k_{DG} - 4 \hat \sigma^k_{DH}, 
\label{Xsecparton2}
\end{align}
and others are the same as those appearing in (\ref{Xsec2}). 
Although $\hat \sigma^k_{DF}=0$ as shown in (\ref{sigma2}), $\hat \sigma^k_{DF1}$ term appears
in (\ref{result}) 
due to the relations (\ref{eom1}), (\ref{rel1}) and (\ref{rel2}).

To write down the partonic hard cross sections in (\ref{result}), 
we further take into account the following relations: 
\beq
&& \hat \sigma^k_{O2}= \hat \sigma^k_{O3},  
\label{O2O3}\\[5pt]
&& \hat \sigma^k_{DF1}=-{1\over 2} \hat \sigma^k_{N1},
\label{DF1N1}\\[5pt]
&& \hat \sigma^k_{DN1} = \hat \sigma^k_{DN2}=\hat \sigma^k_{DO1}= \hat \sigma^k_{DO2}, 
\label{DN12DO12}\\[5pt]
&& \hat \sigma^k_{DN3}= - \hat \sigma^k_{DO3}. 
\label{DN3DO3}
\eeq
The relations (\ref{O2O3}) and (\ref{DF1N1}) are obvious from 
(\ref{sigmaO2}), (\ref{sigmaO3}), (\ref{sigmaN1}) and (\ref{Xsecparton2}), 
and (\ref{DN12DO12}) and (\ref{DN3DO3}) are obtained by explicit calculation of the LO diagrams.  

Then the independent hard cross sections
are given as follows.  

\beq
\begin{dcases}
\hat \sigma^1_{G}= C_F
\frac{2 Q^2}{q_T^3}
\frac{(-1 + \hat z)^2 (-1 -\hat x^2 + \hat z^2 (1-6 \hat x + 6 \hat x^2))}{ \hat x \hat z^3},\\[7pt]
\hat \sigma^2_{G}= C_F
\frac{8}{q_T}
{\hat x (-1 + \hat z)}, \\[7pt]
\hat \sigma^3_{G}= C_F
\frac{2 Q}{q_T^2}
\frac{(-1 +\hat z)  ( \hat z - \hat x -2 \hat z \hat x + \hat z^2(-2 + 4 \hat x))}{\hat z^2},\\[7pt]
\hat \sigma^4_{G}= \frac{1}{2}\hat \sigma^2_G,
\\[7pt]
\hat \sigma^8_{G}= C_F
\frac{2Q}{q_T^2}
\frac{(-1+ \hat z) (- \hat x + \hat z(-1 + 2 \hat x))}{\hat z^2},\\[7pt]
\hat \sigma^9_{G}= C_F
\frac{4}{q_T}
\frac{\hat x (-1 + \hat z)}{\hat z},
\end{dcases}
\label{sigmaG}
\eeq
\beq
\begin{dcases}
\hat \sigma^1_{H}= -C_F
\frac{4Q^2}{q_T^3}
\frac{(-1 + \hat z)^2}{\hat z^2}, \\[7pt]
\hat \sigma^2_{H}= 0,\\[7pt]
\hat \sigma^3_{H}= -C_F
\frac{2Q}{q_T^2}
\frac{(-1+\hat z)(-1 + 2 \hat z)}{\hat z^2},\\[7pt]
\hat \sigma^4_{H}= -C_F
\frac{4}{q_T}
\frac{(-1 + \hat z)}{\hat z},\\[7pt]
\hat \sigma^8_{H}= \hat \sigma^3_H,\\[7pt]
\hat \sigma^9_{H}= \hat \sigma^4_H,
\end{dcases}
\eeq
\beq
\begin{dcases}
\hat \sigma^1_{N1}= C_F
\frac{8 Q^2}{q_T^3}
\frac{ (-1+ \hat z)^2 (3 \hat z (1-2\hat x)\hat x + \hat x (1+ \hat x) + 
\hat z^2(1-6\hat x + 6\hat x^2))}{\hat x \hat z^3 },\\[7pt]
\hat \sigma^2_{N1}= C_F
\frac{32}{q_T}
\frac{\hat x (-1+\hat z)^2}{\hat z}, \\[7pt]
\hat \sigma^3_{N1}= C_F
\frac{8Q}{q_T^2}
\frac{ (-1 + \hat z)^2(-1 -2\hat x + \hat z(-2 + 4 \hat x))}{\hat z^2},\\[7pt]
\hat \sigma^4_{N1}= -C_F
\frac{8}{q_T}
\frac{(-1+\hat z)(1 -2(-1 + \hat z)\hat x)}{\hat z}, \\[7pt]
\hat \sigma^8_{N1}= -C_F
\frac{8Q}{q_T^2}
\frac{ (-1 + \hat z)^2}{\hat z^2},\\[7pt]
\hat \sigma^9_{N1}=- C_F
\frac{8}{q_T}
\frac{(-1+ \hat z)}{\hat z},
\end{dcases}
\eeq
\beq
\begin{dcases}
\hat \sigma^1_{N2}=- C_F
\frac{4Q^2}{q_T^3}
\frac{(-1+ \hat z)^2 (-1+(-1 + 3 \hat z)\hat x)}{\hat z^3 },\\[7pt]
\hat \sigma^2_{N2}=- C_F
\frac{8}{q_T}
\frac{\hat x (-1+\hat z)}{\hat z }, \\[7pt]
\hat \sigma^3_{N2}=- C_F
\frac{2Q}{q_T^2}
\frac{(-1+\hat z)(-3(1+ \hat x)+ \hat z (3+4 \hat x))}{\hat z^2 },\\[7pt]
\hat \sigma^4_{N2}= -C_F
\frac{4}{q_T}
\frac{(-1+\hat z)(2+ \hat x)}{\hat z},\\[7pt]
\hat \sigma^8_{N2}= -C_F
\frac{2Q}{q_T^2}
\frac{(-1+\hat z) (-3 - \hat x + \hat z(3 + 2 \hat x))}{\hat z^2},\\[7pt]
\hat \sigma^9_{N2}= \hat \sigma^4_{N2},
\end{dcases} 
\eeq
\beq
\begin{dcases}
\hat \sigma^1_{N3}= -C_F
\frac{4 Q^2}{q_T^3}
\frac{(-1+\hat z)^2(3 \hat z (2- 3 \hat x)\hat x + \hat x (1+ \hat x) + 2\hat z^2(1-6\hat x + 6 \hat x^2))}{\hat x \hat z^3 },\\[7pt]
\hat \sigma^2_{N3}= -C_F
\frac{8}{q_T}
\frac{\hat x (-1+\hat z)(-3 + 4 \hat z)}{\hat z},\\[7pt]
\hat \sigma^3_{N3}= -C_F
\frac{2 Q}{q_T^2}
\frac{(-1+\hat z)(1 + \hat z(7 - 20 \hat x) + 5\hat x + 8\hat z^2(-1 + 2\hat x))}{\hat z^2 },\\[7pt]
\hat \sigma^4_{N3}= \frac{1}{2}\hat \sigma^2_{N3},\\[7pt]
\hat \sigma^8_{N3}= -C_F
\frac{2Q}{q_T^2}
\frac{(-1+\hat z)(1 - \hat x + \hat z(-1 + 2 \hat x))}{\hat z^2},\\[7pt]
\hat \sigma^9_{N3}= -C_F
\frac{4}{q_T}
\frac{\hat x (-1+\hat z) }{\hat z},
\end{dcases} 
\eeq
\beq
\begin{dcases}
\hat \sigma^1_{DN1}= C_F
\frac{2Q^2}{q_T^3}
\frac{(-1+\hat z)^2(2 \hat z (1-3\hat x)\hat x + (1+\hat x)^2 + \hat z^2(1-6\hat x +6\hat x^2))}{\hat x \hat z^3 },\\[7pt]
\hat \sigma^2_{DN1}= \frac{1}{4}\hat \sigma^2_{N1}, \\[7pt]
\hat \sigma^3_{DN1}= C_F
\frac{4Q}{q_T^2}
\frac{(-1+\hat z)^2(-1 -\hat x + \hat z(-1 + 2\hat x))}{\hat z^2 },\\[7pt]
\hat \sigma^4_{DN1}= -C_F
\frac{4}{q_T}
\frac{(-1+\hat z)(1 + \hat x - \hat x \hat z)}{\hat z},\\[7pt]
\hat \sigma^8_{DN1}= \frac{1}{2}\hat \sigma^8_{N1},\\[7pt]
\hat \sigma^9_{DN1}= \frac{1}{2}\hat \sigma^9_{N1},
\end{dcases} 
\eeq
\beq
\begin{dcases}
\hat \sigma^1_{DN3}=- C_F
\frac{4Q^2}{q_T^3}
\frac{(-1+\hat z)^2(1 + 2 \hat z(2-3 \hat x)\hat x + \hat x^2 + \hat z^2(1-6\hat x + 6\hat x^2))}{\hat x \hat z^3 },\\[7pt]
\hat \sigma^2_{DN3}= -2 \hat \sigma^2_{DN1},\\[7pt]
\hat \sigma^3_{DN3}= -C_F
\frac{8Q}{q_T^2}
\frac{(-1+\hat z)^2( -\hat x + \hat z(-1 + 2\hat x))}{\hat z^2 },\\[7pt]
\hat \sigma^4_{DN3}= - \hat \sigma^2_{DN1},\\[7pt]
\hat \sigma^8_{DN3}= 0,\\[7pt]
\hat \sigma^9_{DN3}=0,
\end{dcases} 
\eeq
\beq
\begin{dcases}
\hat \sigma^1_{O1}= -C_F
\frac{8Q^2}{q_T^3}
\frac{ (-1+\hat z)^2(-1-\hat x + \hat z(-2 + 3 \hat x))}{\hat z^3 }, \\[7pt]
\hat \sigma^2_{O1}= - C_F
\frac{16}{q_T}
\frac{ \hat x (-1+\hat z)}{\hat z}, \\[7pt]
\hat \sigma^3_{O1}= -C_F
\frac{4Q}{q_T^2}
\frac{ (-1+\hat z)(-1 -3\hat x + \hat z(-1 + 4\hat x))}{\hat z^2 }, \\[7pt]
\hat \sigma^4_{O1}= \frac{1}{2}\hat \sigma^2_{O1}, \\[7pt]
\hat \sigma^8_{O1}= - C_F
\frac{4Q}{q_T^2}
\frac{(-1+\hat z) (-1 -\hat x + \hat z(-1 + 2 \hat x))}{\hat z^2 }, \\[7pt]
\hat \sigma^9_{O1}= \frac{1}{2}\hat \sigma^2_{O1},
\end{dcases} 
\eeq
\beq
\begin{dcases}
\hat \sigma^1_{O2}= C_F
\frac{4Q^2}{q_T^3}
\frac{ (-1 + \hat z)^2 (1 + \hat x + 3\hat z(2 - 3 \hat x)\hat x + 2 \hat x^2 + \hat z^2(1-6 \hat x + 6\hat x^2))}{\hat x \hat z^3 }, \\[7pt]
\hat \sigma^2_{O2}= - C_F
\frac{8}{q_T}
\frac{\hat x (-1 +\hat z)(3-2 \hat z)}{\hat z}, \\[7pt]
\hat \sigma^3_{O2}= C_F
\frac{2Q}{q_T^2}
\frac{(-1+ \hat z)(1 + \hat z(5-16 \hat x)+ 7 \hat x + \hat z^2(-4 + 8 \hat x))}{\hat z^2 }, \\[7pt]
\hat \sigma^4_{O2}= \frac{1}{2}\hat \sigma^2_{O2}, \\[7pt]
\hat \sigma^8_{O2}= \frac{1}{2} \hat \sigma^8_{O1}, \\[7pt]
\hat \sigma^9_{O2}= \frac{1}{2} \hat \sigma^9_{O1},
\end{dcases} 
\eeq
\beq
\begin{dcases}
\hat \sigma^1_{DF2}= \frac{C_F}{N}
\frac{Q^2}{q_T^3}
\frac{(-1+\hat z)(1 + \hat x - \hat z \hat x)}{\hat x \hat z^2 }, \\[7pt]
\hat \sigma^2_{DF2}= 0, \\[7pt]
\hat \sigma^3_{DF2}= - \frac{C_F}{N}
\frac{Q}{q_T^2}
\frac{(-1 + \hat z)}{\hat z}, \\[7pt]
\hat \sigma^4_{DF2}= -\frac{C_F}{N}
\frac{1}{q_T}, \\[7pt]
\hat \sigma^8_{DF2}= \hat \sigma^3_{DF2} ,\\[7pt]
\hat \sigma^9_{DF2}= \hat \sigma^4_{DF2}, 
\end{dcases} 
\eeq
\beq
\begin{dcases}
\hat \sigma^1_{DF3}= -\frac{C_F}{N}
\frac{1}{q_T}
\frac{3\hat z(1-2\hat x)\hat x + \hat x (1+\hat x) + \hat z^2(1-6\hat x +6\hat x^2)}{\hat z^2}, \\[7pt]
\hat \sigma^2_{DF3}= -\frac{C_F}{N}
\frac{4 q_T }{Q^2}
\hat x^2 , \\[7pt]
\hat \sigma^3_{DF3}= -\frac{C_F}{N}
\frac{1}{Q}
\frac{\hat x(-1 -2 \hat x + \hat z(-2 + 4 \hat x))}{\hat z }, \\[7pt]
\hat \sigma^4_{DF3}= -\frac{C_F}{N}
\frac{1}{q_T}
\frac{(-1+ \hat x) (-1 + 2(-1 + \hat z)\hat x)}{\hat z}, \\[7pt]
\hat \sigma^8_{DF3}= \frac{C_F}{N}
\frac{1}{Q}
\frac{\hat x}{\hat z }, \\[7pt]
\hat \sigma^9_{DF3}= \frac{C_F}{N}
\frac{1}{q_T}
\frac{-1 + \hat x}{\hat z},
\end{dcases} 
\eeq
\beq
\begin{dcases}
 \hat \sigma^1_{DF4}= 
 \frac{C_F}{N}
\frac{Q^2}{q_T^3}
\frac{(-1+\hat z)(-1 + \hat z + 5 \hat x -6 \hat z \hat x -6 \hat x^2 + 6 \hat z \hat x^2)}{\hat x^2} \\
\ \  +\frac{C_F}{q_T}
\frac{(\hat x-1)^2(1+\hat x) - \hat z (\hat x-1)^2
(1+6 \hat x) - \hat z^3(1-6\hat x + 6\hat x^2) + \hat z^2(1+\hat x - 6\hat x^2 + 
6\hat x^3)}
{\hat z \hat x(-1 + \hat z + \hat x)},\\[7pt]
\hat \sigma^2_{DF4}= \frac{C_F}{N}
\frac{4}{q_T}
{\hat z (-1 + \hat z)}
-C_F
\frac{4q_T}{Q^2}
\frac{\hat z (1 + \hat z - \hat x)\hat x}{(-1 + \hat z + \hat x)},
\\[7pt]
 \hat \sigma^3_{DF4}= \frac{C_F}{N}
\frac{2Q}{q_T^2}
\frac{(-1+\hat z) (-\hat x + \hat z(-1 + 2\hat x))}{ \hat x}
 +
\frac{2C_F}{Q}
\frac{(\hat z +\hat z^2 + \hat x - 2\hat z \hat x -2 \hat z^2 \hat x - 
\hat x^2 + 2 \hat z \hat x^2)}{(-1 + \hat z + \hat x)},\ \\[7pt]
\hat \sigma^4_{DF4}= \frac{C_F}{N}
\frac{1}{q_T}
\frac{(1 + \hat x + 2\hat z^2 \hat x - \hat z(1+2 \hat x))}{\hat x}
-C_F
\frac{2}{q_T}
\frac{(-1 + \hat x)(1 + \hat z^2 \hat x + \hat x^2 - 
\hat z(1+ \hat x^2))}{\hat x(-1 + \hat z +\hat x)},\\[7pt]
\hat \sigma^8_{DF4}= \frac{C_F}{N}
\frac{2 }{Q} \hat z
-C_F
\frac{4}{Q}
\frac{ \hat z (-1 + \hat x)}{(-1 + \hat z + \hat x)},\\[7pt]
\hat \sigma^9_{DF4}= \frac{C_F}{N}
\frac{1}{q_T}
\frac{ (1 - \hat x + \hat z( -1 + 2\hat x))}{\hat x}
- C_F
\frac{2}{q_T}
\frac{(-1+ \hat x)(1- \hat x + \hat z(-1 + 2 \hat x))}{\hat x (-1 + \hat z + \hat x)},
\end{dcases}
\eeq
\beq
\begin{dcases}
\hat \sigma^1_{DF5}= \frac{C_F}{N}
\frac{1}{q_T}
\frac{(-1 + \hat x)^2(1 + \hat x + 6 \hat z^2 \hat x - \hat z(1 + 6 \hat x))}{ \hat z^2 \hat x} \\
\quad -
\frac{C_F}{q_T}
\frac{(\hat x-1)^2(1+\hat x) - \hat z (\hat x-1)^2(1+6 \hat x)
-\hat z^3(1-6\hat x + 6\hat x^2) + \hat z^2(1+\hat x - 6\hat x^2 + 6\hat x^3)}{
 \hat z \hat x(-1 + \hat z + \hat x)}, \\[7pt]
\hat \sigma^2_{DF5}= \frac{C_F}{N}
\frac{4q_T}{Q^2}
{ (-1 + \hat x) \hat x}
+C_F
\frac{4 q_T}{Q^2}
\frac{\hat z (1 + \hat z - \hat x)\hat x}{(-1 + \hat z + \hat x)}, \\[7pt]
 \hat \sigma^3_{DF5}= \frac{C_F}{N}
\frac{2}{Q}
\frac{(-1 + \hat x) (-\hat x + \hat z(-1 + 2\hat x))}{\hat z }
-\frac{2C_F}{Q}
\frac{(\hat z +\hat z^2 + \hat x - 2\hat z \hat x -2 \hat z^2 \hat x - \hat x^2 + 2 \hat z \hat x^2)}{(-1 + \hat z + \hat x)},\\[7pt]
\hat \sigma^4_{DF5}= \frac{C_F}{N}
\frac{1}{q_T}
\frac{ (-1+ \hat x)(-1 + \hat z + \hat x - 2\hat z \hat x -2 \hat x^2 + 2 \hat z \hat x^2 )}{\hat x \hat z }\\
\qquad\quad+C_F
\frac{2}{q_T}
\frac{(-1+ \hat x) (1 + \hat z^2 \hat x + \hat x^2 - \hat z(1+ \hat x^2))}{\hat x (-1 + \hat z +\hat x)},\\[7pt]
   \hat \sigma^8_{DF5}= - \frac{C_F}{N}
\frac{2}{Q}
{(-1 + \hat x)}
+C_F
\frac{4}{Q}
\frac{\hat z (-1 + \hat x)}{(-1 + \hat z + \hat x)},\\[7pt]
  \hat \sigma^9_{DF5}= -\frac{C_F}{N}
\frac{1}{q_T}
\frac{ (-1 + \hat x)(1 - \hat x + \hat z( -1 + 2\hat x))}{\hat x \hat z}
+ C_F
\frac{2}{q_T}
\frac{(-1 + \hat x) (1- \hat x + \hat z(-1 + 2 \hat x))}{\hat x (-1 + \hat z + \hat x)},
\end{dcases}
\eeq
\beq
\begin{dcases}
\hat \sigma^1_{GF2}= -\frac{C_F}{N},
\frac{Q^2}{q_T^3}
\frac{(-1+\hat z)(1 - \hat x + \hat z \hat x)}{\hat x \hat z^2 }, \\[7pt]
\hat \sigma^2_{GF2}= 0, \\[7pt]
\hat \sigma^3_{GF2}= \hat \sigma^3_{DF2}, \\[7pt]
\hat \sigma^4_{GF2}=\hat \sigma^4_{DF2}, \\[7pt]
\hat \sigma^8_{GF2}= \hat \sigma^8_{DF2}, \\[7pt]
\hat \sigma^9_{GF2}= \hat \sigma^9_{DF2},
\end{dcases} 
\eeq
\beq
\begin{dcases}
\hat \sigma^1_{GF3}= -\frac{C_F}{N}
\frac{1}{q_T}
\frac{\hat z (5-6 \hat x)\hat x + \hat x (-1+\hat x)+\hat z^2(1-6\hat x + 6\hat x^2)}{\hat z^2}, \\[7pt]
\hat \sigma^2_{GF3}= \hat \sigma^2_{DF3}, \\[7pt]
\hat \sigma^3_{GF3}= -\frac{C_F}{N}
\frac{1}{Q}
\frac{\hat x(-1+2\hat z)(-1+2\hat x)}{\hat z }, \\[7pt]
\hat \sigma^4_{GF3}= -\frac{C_F}{N}
\frac{1}{q_T}
\frac{(-1+ \hat x) (1 + 2(-1 + \hat z)\hat x)}{\hat z}, \\[7pt]
\hat \sigma^8_{GF3}= - \hat \sigma^8_{DF3}, \\[7pt]
\hat \sigma^9_{GF3}=- \hat \sigma^9_{DF3},
\end{dcases}
\eeq 
\beq
\begin{dcases}
\hat \sigma^1_{GF4}= \frac{C_F}{N}
\frac{Q^2}{q_T^3}
\frac{(-1+\hat z) (-1 + \hat z + 7 \hat x -6 \hat z \hat x -6 \hat x^2 + 6 \hat z \hat x^2)}{\hat x^2 } \\
\qquad\qquad-C_F
\frac{1}{q_T}
\frac{(-1+\hat x)^2 -6\hat z(-1+\hat x)\hat x + \hat z^2(1 - 6\hat x + 6\hat x^2)}{ \hat z \hat x},\\[7pt]
\hat \sigma^2_{GF4}= \frac{C_F}{N}
\frac{4}{q_T}
{ \hat z (-1+\hat z)}
-C_F
\frac{4q_T}{Q^2}
{ \hat z  \hat x},\\[7pt]
 \hat \sigma^3_{GF4}= \frac{C_F}{N}
\frac{2Q}{q_T^2}
\frac{(-1+\hat z) (1 -\hat x + \hat z(-1 + 2\hat x))}{ \hat x}
-C_F
\frac{2}{Q}
{(1-\hat x + \hat z(-1+2\hat x))},\\[7pt]
  \hat \sigma^4_{GF4}= \frac{C_F}{N}
\frac{1}{q_T}
\frac{ (-1 +\hat z + \hat x -2\hat z \hat x + 2\hat z^2 \hat x)}{\hat x}
-C_F
\frac{2}{q_T}
\frac{ (-1+\hat x) (1+ (-1+\hat z)\hat x) }{\hat x},\\[7pt]
   \hat \sigma^8_{GF4}= \frac{C_F}{N}
\frac{2Q}{q_T^2}
{(-1+\hat z)}
-C_F
\frac{2 }{Q},\\[7pt]
  \hat \sigma^9_{GF4}= \frac{C_F}{N}
\frac{1}{q_T}
\frac{ (-1+\hat z - \hat x + 2\hat z \hat x)}{\hat x}
- C_F
\frac{2}{q_T}
\frac{(-1 +\hat x)}{\hat x},
\end{dcases}
\eeq
\beq
\begin{dcases}
 \hat \sigma^1_{GF5}= \frac{C_F}{N}
\frac{1}{q_T}
\frac{(-1 +\hat x)^2 (-1 + \hat z+\hat x  -6 \hat z \hat x +6\hat z^2 \hat x)}{ \hat z^2 \hat x} \\
\qquad\qquad -C_F
\frac{1}{q_T}
\frac{(-1+\hat x)^2 -6\hat z(-1+\hat x)\hat x + \hat z^2(1 - 6\hat x + 6\hat x^2)}{\hat z \hat x},\\[7pt]
  \hat \sigma^2_{GF5}= \frac{C_F}{N}
\frac{4q_T}{Q^2}
{(-1+\hat x) \hat x}
-C_F
\frac{4q_T}{Q^2}
{ \hat z  \hat x},\\[7pt]
 \hat \sigma^3_{GF5}=\frac{C_F}{N}
\frac{2}{Q}
\frac{(-1+\hat x)(1 -\hat x + \hat z(-1 + 2\hat x))}{ \hat z}
-C_F
\frac{2}{Q}
{(1-\hat x + \hat z(-1+2\hat x))},\\[7pt]
  \hat \sigma^4_{GF5}=\frac{C_F}{N}
\frac{1}{q_T}
\frac{ (-1+\hat x)(-1 +\hat z +3 \hat x -2\hat z \hat x -2\hat x^2 + 2\hat z \hat x^2)}{\hat x \hat z} \\
\qquad\qquad -C_F
\frac{2}{q_T}
\frac{ (-1+\hat x) (1+ (-1+\hat z)\hat x) }{\hat x},\\[7pt]
   \hat \sigma^8_{GF5}= - \frac{C_F}{N}
\frac{2}{Q}
\frac{(-1+\hat z)(-1+\hat x)}{\hat z}
-C_F
\frac{2 }{Q},\\[7pt]
  \hat \sigma^9_{GF5}= - \frac{C_F}{N}
\frac{1}{q_T}
\frac{(-1+\hat x) (1-3\hat x +\hat z(-1 + 2\hat x))}{\hat x \hat z}
- C_F
\frac{2}{q_T}
\frac{(-1+\hat x)}{\hat x}.
\end{dcases}
\label{sigmaGF5}
\eeq
Eqs. (\ref{sigmaG})-(\ref{sigmaGF5}) and the relations (\ref{O2O3}), 
(\ref{DF1N1}), 
(\ref{DN12DO12}),
(\ref{DN3DO3}) specify all the partonic cross sections in the final
formula (\ref{result}).

\section{Summary}

In this paper we have studied the transversely polarized spin-1/2 hyperon production
in SIDIS, $ep\to e\Lambda^\uparrow X$.  Specifically,
we have derived the LO twist-3 gluon FF contribution
to the polarized cross section.  Since the twist-3 gluon FFs are related to
the $q\bar{q}g$-FFs through the EOM relations and the LIRs, we consistently took
into account the latter contribution together.  
This has completed the twist-3 LO cross section for this process
together with the results for the contribution from the twist-3 DF and the
twist-3 quark FFs derived in \cite{Koike:2022ddx}.  
The final result for the cross section is given in
(\ref{result}).  It consists of five components with different azimuthal structures as
\begin{align}
\frac{\drm^6 \sigma}{\drm x_{bj} \drm Q^2 \drm z_f \drm q_T^2 \drm \phi \drm \chi} =
&\mathcal{F}_1 \sin \Phi_S + \mathcal{F}_2 \sin \Phi_S 
\cos \varphi + \mathcal{F}_3 \sin \Phi_S \cos 2\varphi {\nonumber}\\
&+\mathcal{F}_4 \cos \Phi_S \sin \varphi + \mathcal{F}_5 \cos \Phi_S \sin 2\varphi, 
\end{align}
where $\varphi=\phi-\chi$ is the relative azimuthal angle between the lepton ($\phi$) and the
hadron ($\chi$) planes and $\Phi_S$ is the azimuthal angle of
the transverse spin vector of $\Lambda^\uparrow$ measured from the hadron plane
with the structure functions 
$\mathcal{F}_{1,2,3,4,5}$ written as convolution of the twist-3 FFs and the quark DF
in the proton and the partonic hard cross sections. 
The LO cross section given in \cite{Koike:2022ddx} 
and the present study contains several unknown nonperturbative functions, and their
determination requires global anayses of many twist-3 processes in which the same twist-3 functions
appear.  Information from analyses of small-$P_T$ data in terms of the TMD factorization is
also of great help to constrain some of the twist-3 functions.  
In any case our twist-3 cross section formula
is the starting point of analyzing the large-$P_T$ hyperon polarization in SIDIS 
which we hope to
be measured 
in the future EIC experiment.

\section*{Acknowledgments}
This work has been supported by 
the establishment of Niigata university fellowships towards the creation of science
technology innovation (R.I.),
the Grant-in-Aid for
Scientific Research from the Japanese Society of Promotion of Science
under Contract Nos.~19K03843 (Y.K.) and 18J11148 (K.Y.),
National Natural Science Foundation in China 
under grant No. 11950410495, Guangdong Natural Science Foundation under
No. 2020A1515010794
and research startup funding at South China
Normal University (S.Y.).

\bibliographystyle{unsrt} 
\bibliography{ref2}

\end{document}